\documentclass[%
 aps,
 amsmath,amssymb,
 reprint,%
superscriptaddress,
longbibliography
]{revtex4-1}

\usepackage{float}
\usepackage{graphicx}
\usepackage{subfigure}
\usepackage{array}
\usepackage{amsmath}
\usepackage{color}
\usepackage{bm}
\usepackage{soul}
\usepackage[normalem]{ulem}
\usepackage{mathtools}

\makeatletter
\newcommand*{\balancecolsandclearpage}{%
  \close@column@grid
  \cleardoublepage
  \twocolumngrid
}
\makeatother

\usepackage{xspace}
\definecolor{cola}{rgb}{0.7,0.1,0.1}
\definecolor{colb}{rgb}{0.9,0.4,0}
\definecolor{colc}{rgb}{0.3,0.7,0}
\definecolor{cold}{rgb}{0,0.35,0.75}
\definecolor{cole}{rgb}{0.63, 0.13, 0.94}

\newcommand{\Tone}[0]{{$T_1$}\xspace}

\newcommand{\ket}[1]{\left|#1\right\rangle}

\newcommand{\bra}[1]{\left\langle#1\right|}

\graphicspath{{figures/}}

\begin{document}

\title{Longitudinal spin-relaxation of donor-bound electrons \\ in direct bandgap semiconductors}

\author{Todd~Karin}
\thanks{These authors contributed equally to this work.}
\affiliation{Department of Physics, University of Washington, Seattle, Washington 98195, USA}
\author{Xiayu~Linpeng}
\thanks{These authors contributed equally to this work.}
\affiliation{Department of Physics, University of Washington, Seattle, Washington 98195, USA}
\author{M. V.~Durnev}
\affiliation{Ioffe Institute, 194021 St.-Petersburg, Russia}
\author{Russell~Barbour}
\affiliation{Department of Physics, University of Washington, Seattle, Washington 98195, USA}
\author{M. M.~Glazov}
\affiliation{Ioffe Institute, 194021 St.-Petersburg, Russia}
\author{E. Ya. Sherman}
\affiliation{Department of Physical Chemistry, The University of the Basque Country, 48080 Bilbao, Spain}
\affiliation{IKERBASQUE Basque Foundation for Science, Bilbao, Spain}
\author{Simon~Watkins}
\affiliation{Department of Physics, Simon Fraser University, Burnaby, BC V5A-1S6, Canada}
\author{Satoru~Seto}
\affiliation{National Institute of Technology, Ishikawa College, Tsubata, Kahoku, Ishikawa 929-0392, Japan}
\author{Kai-Mei~C.~Fu}
\affiliation{Department of Physics, University of Washington, Seattle, Washington 98195, USA}
\affiliation{Department of Electrical Engineering, University of Washington, Seattle, Washington 98195, USA}

\begin{abstract}
We measure the donor-bound electron longitudinal spin-relaxation time ($T_1$) as a function of magnetic field ($B$) in three high-purity direct-bandgap semiconductors: GaAs, InP, and CdTe,  observing a maximum $T_1$ of 1.4~ms, 0.4~ms and 1.2~ms, respectively. In GaAs and InP at low  magnetic field, up to $\sim$2~T, the spin-relaxation mechanism is strongly density and temperature dependent and is attributed to the random precession of the electron spin in hyperfine fields caused by the lattice nuclear spins. In all three semiconductors at high magnetic field, we observe a power-law dependence ${T_1 \propto B^{-\nu}}$ with ${3\lesssim \nu \lesssim 4}$. Our theory predicts that the direct spin-phonon interaction is important in all three materials in this regime in contrast to quantum dot structures. In addition, the ``admixture'' mechanism caused by Dresselhaus spin-orbit coupling combined with single-phonon processes has a comparable contribution in GaAs. We find excellent agreement between high-field theory and experiment for GaAs and CdTe with no free parameters, however a significant discrepancy exists for InP. 
\end{abstract}

\maketitle

\section{Introduction}

In the last decade, the prospects for spin-based quantum information have spurred renewed interest in the fundamental mechanisms for spin relaxation in semiconductors~\cite{ref:jarmola2012tmf, toddkarin:Tyryshkin2006, ref:tribollet2009ten,toddkarin:Khaetskii2001}. Shallow impurities in direct-bandgap materials are promising candidates for quantum applications relying on spin-photon interfaces~\cite{ref:barrett2005ehf, ref:childress2005ftq, ref:simon2010qm}, as these systems boast high optical homogeneity~\cite{toddkarin:Fu2005}, strong spin-photon coupling, and the potential in II-VI materials~\cite{toddkarin:DeGreve2010} to enhance spin coherence times with isotope purification~\cite{toddkarin:Balasubramanian2009, toddkarin:Tyryshkin2012}. While electron spin relaxation is
now relatively well understood in III-V semiconductor quantum dots both theoretically and experimentally~\cite{toddkarin:Khaetskii2001,toddkarin:Merkulov2002,toddkarin:Kroutvar2004,Elzerman:SR:2004,toddkarin:Heiss2005,toddkarin:Amasha2008,toddkarin:Lu2010}, it is still an open question whether the same processes dominate in the similar direct band-gap donor system. In contrast to quantum dots, in which the size, shape, composition, and strain field for each dot are to a large extent unknown, the physical properties relevant to spin relaxation for the homogeneous donor system have been measured. This enables quantitative comparison of spin-relaxation rates between theory and experiment which should help predict which donor systems are most promising for future applications. 


Here we measure the longitudinal spin-flip time \Tone, the fundamental limit for the storage time for quantum information, in three semiconductors: GaAs, InP, and CdTe. All three are direct bandgap materials with similar band structure allowing for the optical pumping of the donor-bound electron spins under resonant exciton excitation.
We show that at low magnetic fields, ${T_1}$ is proportional to ${B^2}$ with a proportionality constant highly dependent on temperature and donor density. At high magnetic fields, we find that \Tone is proportional to $B^{-\nu}$, with the power $\nu$ in the range ${3\lesssim \nu \lesssim 4}$. The competition of these two 
dependencies leads to a maximum of \Tone in GaAs and InP at relatively high magnetic field: $(1.4 \pm 0.1)~\mathrm{ms} \mathrm{~at~} 4~\mathrm{T}$ for GaAs and $(0.40 \pm 0.01)~\mathrm{ms} \mathrm{~at~} 1.9~\mathrm{T}$ for InP. Due to technical issues, we are unable to observe this maximum for CdTe; however, the highest \Tone measured is $(1.23\pm 0.07)~\mathrm{ms} \mathrm{~at~} 1.1~\mathrm{T}$ with \Tone~expected to rapidly increase at lower fields. 

The low magnetic-field \Tone behavior for GaAs and InP is consistent with a spin relaxation mechanism controlled by the hyperfine coupling of the electron spin with static fluctuations of the host-lattice nuclear spins. In this situation, spin precession is randomized due to the finite electron correlation time at each donor site~\cite{toddkarin:opticalOrientationBook,toddkarin:Dzhioev2002}. Although the mechanism for the extremely-short correlation time $\tau_c$ ($\tau_{c,\mathrm{GaAs}} \simeq 25~\mathrm{ns}$, $\tau_{c,\mathrm{InP}} \simeq 40~\mathrm{ns}$) is not completely clear, our measurement is consistent with prior works~\cite{toddkarin:Dzhioev2002,toddkarin:Berski2015}. Our results show that 
the nuclear-spin environment, known to be the dominant factor in spin dephasing~\cite{toddkarin:Merkulov2002,toddkarin:Tyryshkin2006}, plays an important role in longitudinal relaxation even at low doping densities ($\sim$10$^{14}$~cm$^{-3}$) and moderate magnetic fields (up to several tesla).

On the high-field side, the similar magnetic-field dependence observed in all three semiconductors is suggestive of a universal mechanism. We theoretically investigate the dominant spin-relaxation mechanisms and find that two mechanisms, (i)  the direct spin-phonon interaction and, (ii) the admixture mechanism caused by Dresselhaus spin-orbit coupling combined with the piezoelectric electron-phonon interaction, can account for the magnitude of the observed relaxation in GaAs and CdTe. The strength of the direct spin-phonon interaction is surprising because it was found to be negligible in the similar quantum dot system~\cite{toddkarin:Khaetskii2001}. 
We find, however, that both interactions are too weak to account for the observed relaxation in InP.

The paper is organized as follows: Section~\ref{sec:samples} presents the studied samples and experimental technique to measure \Tone, the experimental results are summarized in Sec.~\ref{sec:exper}. Section~\ref{sec:theory} presents the theory and comparison with experiment. The paper is summarized by a short conclusion~\ref{sec:conclusion}. Appendices include additional experimental and theoretical details.

\section{Samples and experimental technique}\label{sec:samples}

\begin{figure}[tb]
\includegraphics[width=8.6cm]{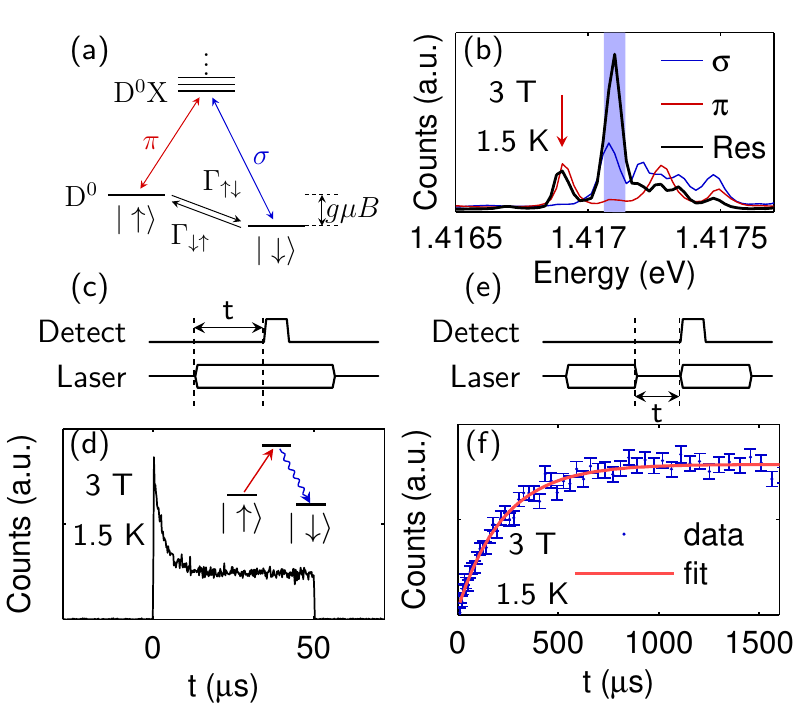}
\caption{
(a) Energy level diagram for the InP donor system.
(b) Photoluminescence spectrum of InP. Excitation at 1.549~eV with 50~$\mu$W power, for the two above-bandgap excitation spectra (red and blue). $\sigma$ ($\pi$) denote linear collection polarization perpendicular (parallel) to the magnetic field. Resonant excitation spectrum (black) uses excitation at 1.417~eV with 100~$\mu$W $\pi$-polarized light, with $\sigma$-polarized light collected.
(c) Pulse sequence for optical pumping. The Ti:Sapphire laser is pulsed on and off repetitively on the \(\pi\) transition, while PL from the \(\sigma\) transition is detected. The time between pulses significantly exceeds \Tone.
(d) Optical pumping trace for InP with laser power 10~$\mu$W. The inset sketches the population transfer process during optical pumping. The amplitude of the exponential curve is proportional to the population in $\uparrow$.
(e) Pulse sequence for \Tone measurement. 
The detector gate-on time is 2~$\mu$s and the laser pulse length is 50~$\mu$s.
(f) \Tone measurement for InP with laser power 10~$\mu$W. The data is fit with an exponential plus a background yielding the time constant ${T_1 = (0.23\pm0.1)~\text{ms}}$. Error bars denote the standard deviation of the recovery signal in each time bin over the many repetitions of the pulse sequence. The corresponding representative data for GaAs and CdTe are given in Appendix \ref{appendix:GaAsCdTe}. All experiments used $\sim$30~$\mu\mathrm{m}$ laser spot size.
 }\label{fig:setup}
\end{figure}

We study two GaAs, three InP, and two CdTe $n$-doped samples with the parameters given in Table~\ref{table:samples}. Spin-relaxation is measured optically in the Voigt geometry (photon wave vector $\mathbf k\perp \mathbf B$) with the magnetic field aligned parallel to the sample surface. Magneto-photoluminescene spectra exhibiting optically resolved Zeeman transitions for all three semiconductors are shown in Appendix~\ref{appendix:magnetoPL}. $\Lambda$-transitions suitable for optically pumping the electron spin are found by resonantly exciting one of the Zeeman sublevels of the neutral donor (D$^0$) to the lowest neutral donor-bound exciton (D$^0$X) transition and observing the corresponding Raman transition. The optically excited and collected transitions for InP (GaAs, CdTe) are labelled in the energy diagram and photoluminescence spectra in Figs.~\ref{fig:setup}(a),(b) [Appendix~\ref{appendix:GaAsCdTe}, Figs.~\ref{fig:exp_CdTe_GaAs}(a),(b),(e),(f)].

 \begin{table}
\caption{Sample parameters. ${N_e= N_D - N_A}$ is the electron density, $\ell$ is the sample thickness. Metal organic vapour phase epitaxy and molecular beam epitaxy are abbreviated by MOCVD and MBE respectively. The InP epilayer is grown directly on an InP substrate.  The GaAs epilayer is grown on 4~microns of Al$_{0.3}$Ga$_{0.7}$As on a GaAs substrate. Further details on sample growth are given in the references.} \label{table:samples}
 \centering
 \begin{tabular}{cccc}
 \hline \hline
 Sample & $N_e$~(cm$^{-3}$)  & $\ell$~($\mu$m) & Growth Method  \\
 \hline 

 InP-1\cite{ref:InPfootnote} & 5.6 $\times 10^{13}$ & 5.1 & MOCVD \\ 
 InP-2\cite{ref:InPfootnote} & 2.3 $\times 10^{14}$ & 7.4 & MOCVD \\ 
 InP-3\cite{ref:InPfootnote} & 1.8 $\times 10^{15}$  & 4.2 &  MOCVD \\ 
 GaAs-1\cite{toddkarin:Stanley1991} & 3$\times 10^{13}$ & 15 & MBE \\ 
 GaAs-2 & 5$\times 10^{13}$ & 10 & MBE \\ 
 CdTe-1\cite{ref:seto1988abb} & $1\times 10^{14}$ & $>$1000 & Bridgman \\
 CdTe-2\cite{ref:CdTefootnote}& $>10^{14}$   & $>$1000 & Bridgman \\
\hline \hline
\end{tabular}
\end{table}

To measure the spin relaxation time in the magnetic field, we optically deplete one of the Zeeman spin sublevels and monitor the recovery of its thermal population in the course of spin relaxation. At high magnetic fields, the optically-resolved spin Raman transitions enable frequency-selective optical pumping of the donor electron state. At low fields, while the transitions cannot be spectrally resolved, optical pumping is still obtained by utilizing the optical polarization selection rules. Optical pumping is confirmed by monitoring the time-dependence of the collected transition intensity during optical excitation after the system has reached thermal equilibrium. A typical high-field optical pumping pulse sequence and photoluminescence trace are depicted in Figs.~\ref{fig:setup}(c),(d). The decrease in photoluminescence intensity is only observed with resonant spin excitation. Two-laser experiments in GaAs have also confirmed that this decrease is due to spin-pumping and not, for example, due to photo-induced ionization~\cite{ref:fu2006msf}. A clear optical pumping signal cannot be observed in the highest purity InP sample, InP-1. The cause is attributed to surface depletion effects discussed further in Appendix~\ref{appendix:depletion}. For the remainder of the paper we will restrict ourselves to the remaining six samples, where reliable signals are detected.

Spin-relaxation measurements are performed by varying the recovery time between optical pumping pulses which are produced by an acousto-optic modulator (AOM) from the output of a narrow-band continuous-wave Ti:Sapphire laser. The AOM extinction ratio, $r_e$ was measured to be ${>}10^4$ giving an upper-bound of the maximum measurable \Tone of $r_e\tau_{op}$, in which $\tau_{op}$ is the characteristic timescale of optical pumping. Given the several microsecond $\tau_{op}$ [Fig.~\ref{fig:setup}(d)], we have the ability to measure \Tone exceeding 10~ms.  The ``Raman'' photoluminescence is collected during the first part of the optical pumping pulse, see Fig.~\ref{fig:setup}(e). As the recovery time increases, we observe an increase in the collected signal as the system returns to thermal equilibrium. At each magnetic field, the recovery is fitted to a weighted exponential with time constant $T_1$~\cite{toddkarin:Turton2003}, as shown in Fig.~\ref{fig:setup}(e). Measurements are performed for fields up to 7.0~T. Reduced visibility of the optical pumping signal places a technical limit on the minimum magnetic field measurement for each sample.

\section{Experimental Results}\label{sec:exper}

\begin{figure}[ht]
\includegraphics[width =\linewidth]{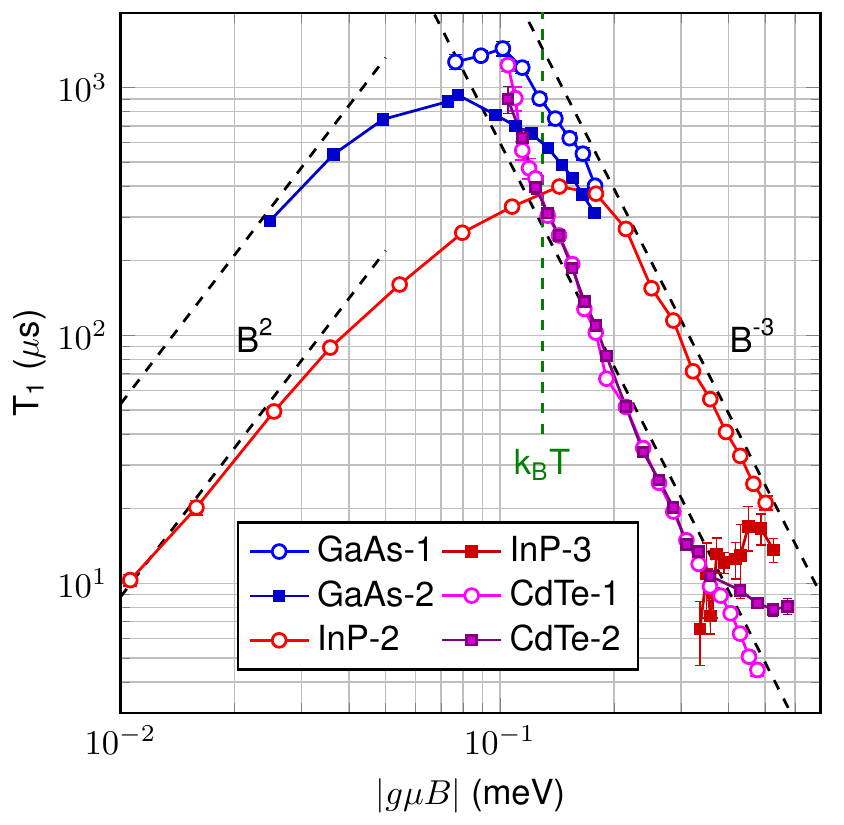}
\caption{\Tone as a function of Zeeman splitting for the six different samples at 1.5~K. The absolute values of the electron $g$-factors used to convert from $B$ to the Zeeman splitting for GaAs, InP and CdTe are 0.44, 1.3, and 1.65, respectively. Sample descriptions are given in Table~\ref{table:samples}. The black dashed lines in the high energy (low energy) side denote a $B^{-3}$ ($B^{2}$) dependence for reference. They are offset from the experimental data for clarity.  The green dashed line denotes the thermal energy ${k_BT}$ for reference.
}
\label{fig:T1expdata}
\end{figure}

\begin{figure}[htb]
    \includegraphics{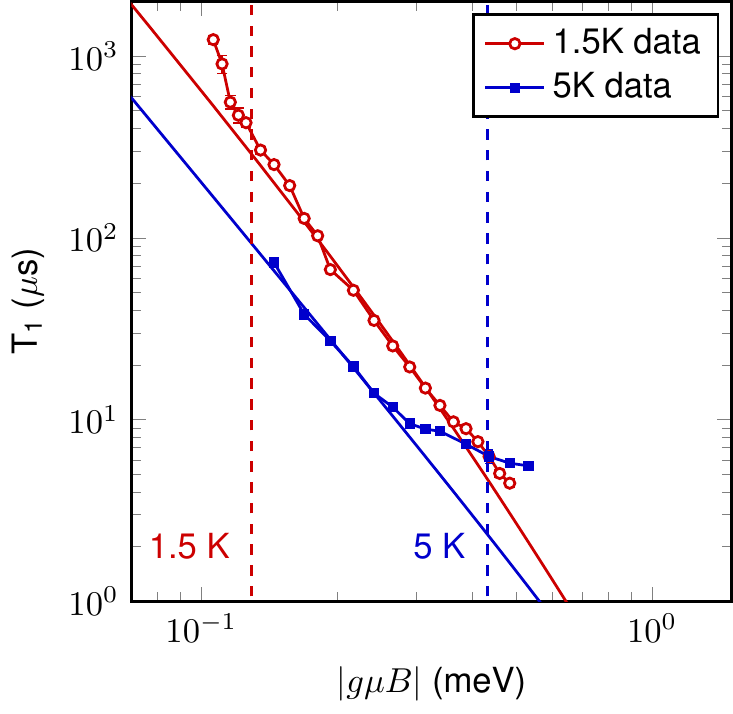}
    \caption{$T_1$ as a function of Zeeman splitting for {CdTe-1} at ${T = 1.5~\text{K}}$ and ${T = 5~\text{K}}$. The red and blue lines are mutually fitted by an empirical formula ${T_1 = b B^4 / F_{ph}}$, where ${b = 2000 ~\mu s/\mathrm{T}^4}$. The red and blue dashed lines denote the energy at 1.5~K and 5~K.}
    \label{fig:tempdataCdTe}
\end{figure}

The longitudinal spin relaxation times $T_1$ as a function of the electron Zeeman splitting ${\Delta E = |g\mu B|}$ for InP, GaAs and CdTe are shown in Fig~\ref{fig:T1expdata}. Here, $g$ is the effective electron $g$-factor and $\mu$ is the Bohr magneton. The data show several notable features. First, all samples approach a \({T_1 \sim B^{-\nu}}\) dependence, with ${3\lesssim \nu \lesssim 4}$, at high magnetic fields. The proportionality constant depends on the semiconductor sample. A $B^{-3}$ dependence, included in Fig.~\ref{fig:T1expdata}, fits all curves well, however we note that higher-field data would be desirable for GaAs because the small electron $g$-factor prevents us from accessing the high-Zeeman-splitting limit, where ${|g\mu B| \gg k_B T}$. Also, a $B^{-4}$ power-law is reasonable for CdTe, as the magnetic field dependence becomes steeper in CdTe with decreasing field, see also Fig.~\ref{fig:tempdataCdTe}. The high-field \(T_1\) process appears to be independent of donor concentration. Even the \Tone curve for the high-density InP-2 sample approaches the InP-1 curve at the highest fields. At low fields, \Tone in InP and GaAs approaches a $B^2$ dependence with a donor-concentration-dependent pre-factor. This is extremely pronounced for the InP samples in which the donor-bound electron density $N_e$, the difference between the donor and acceptor densities in the sample, ${N_D-N_A}$, differs by a factor of 4. The effect is also present in GaAs in which $N_e$  differs by a factor of 1.7. Finally, the maximum $T_1$ observed in all three materials is similar: $T_1=1.4$, 0.4, and 1.2~ms for GaAs, InP, and CdTe respectively. 
\begin{figure}[htb]
    \includegraphics{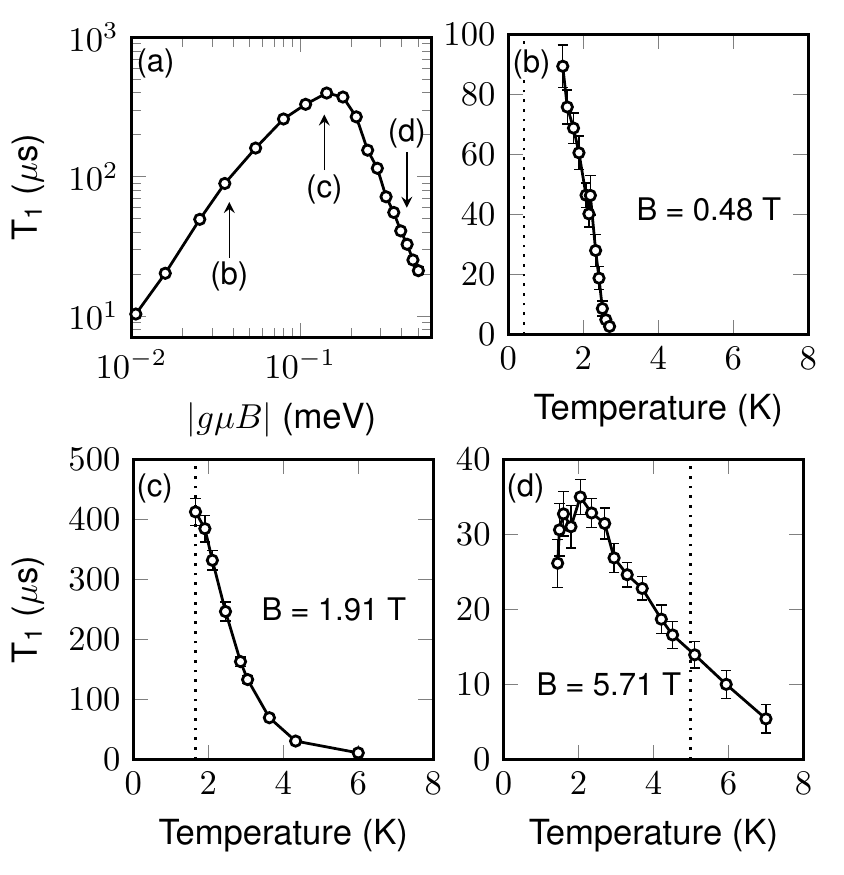}
    \caption{(a) $T_1$ as a function of the Zeeman splitting for {InP-2} at 1.5~K. The arrows show the magnetic field values at which the temperature dependence study was performed. {(b-d)} Temperature dependence of $T_1$ at (b) ${B = 0.48~\text{T}}$, (c) 1.9~T, and (d) 5.7~T. The dotted line denotes ${|g\mu B|/k_B}$. }
    \label{fig:tempdataInP}
\end{figure}

Measurements of the temperature $T$ effect on \Tone are also performed. In InP-2, the sample in which \Tone can be obtained for the largest range of Zeeman energies, $T_1(T)$ was measured at 0.5~T (low field regime), 1.9 T (peak \Tone), and 5.7~T (high-field regime) with the results depicted in Fig.~\ref{fig:tempdataInP}.  In the low-field regime, an extremely-steep inverse dependence of \Tone on temperature is observed indicative of a strong phonon-assisted process. In the high-field regime, the relaxation time is almost independent of temperature at the lowest temperatures in our experiments, and drops with an increase in $T$. This high-field behavior is consistent with a model in which \Tone is inversely dependent on the phonon factor $F_{ph} = 2N_{ph} + 1$, in which ${N_{ph} = [\exp(|g\mu B|/k_BT)-1]^{-1}}$ is the phonon occupation number. A comparison of magnetic-field-dependent measurements at 1.5~K and 5~K for CdTe-1 also support a high-field single-phonon mechanism. The ratio of the two curves in Fig.~\ref{fig:tempdataCdTe} is given by $F_{ph}(5~{\rm K})/F_{ph}(1.5~\rm K)$. 

\section{Theory}\label{sec:theory}

Here we consider the mechanisms resulting in spin relaxation of donor-bound electrons. We start with the limit of relatively-low magnetic fields, where spin relaxation is controlled by the hyperfine coupling of the electron and nuclear spins. Next, we turn to the regime of high enough magnetic fields where the nuclei-induced spin relaxation is unimportant and the spin-flip processes caused by the joint effects of the electron-phonon and the spin-orbit interactions play the major role. 
    
\subsection{Low-field spin-relaxation}\label{sec:lowFieldAdmixture}

\begin{figure}[ht]
    \includegraphics[width=0.9\linewidth]{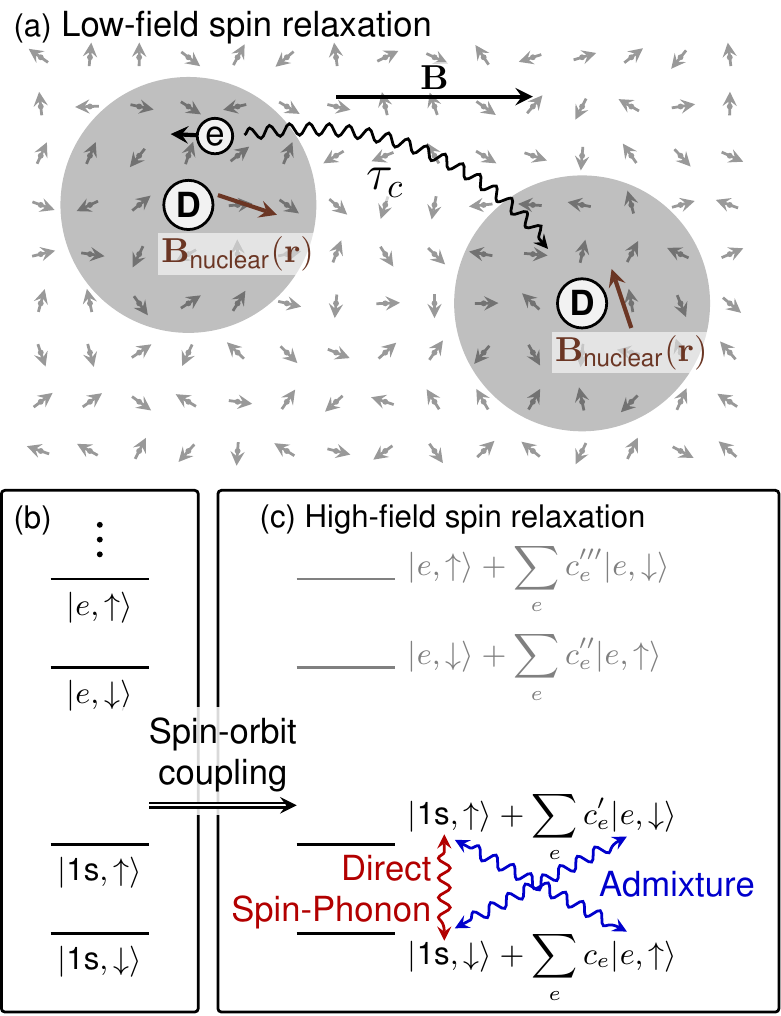}
    \caption{Schematic of spin-relaxation mechanisms.   (a) At low magnetic fields, spin-relaxation is dominated by the interaction of the electron spin with lattice nuclear spins. Panels (b,c) are relevant for the high-field spin relaxation mechanism. (b)~Energy level structure for unperturbed donor-bound electron in magnetic field, described by zero-field quantum numbers.
    (c) Dresselhaus spin-orbit coupling mixes states with opposite spin and different angular momentum components. In the admixture mechanism, phonons cause relaxation between the two eigenstates via the components with like spin. The direct spin-phonon interaction causes spin-relaxation via the components with opposite spin. 
 }
    \label{fig:mechanism}
\end{figure}

At low temperatures and low donor densities, the electrons in bulk semiconductors are localized. At low and moderate magnetic fields, the electron spin relaxation is controlled by the hyperfine interaction with the host lattice nuclei~\cite{toddkarin:opticalOrientationBook,toddkarin:Dyakonov2008}. The spin dynamics of the electron in the ensemble of donors obey the set of kinetic equations~\cite{PhysRevB.91.195301,PhysRevB.92.014206}
\begin{equation}
\label{kinetic}
\frac{d \bm S_i}{dt} + \bm S_i \times \bm \Omega_i = \bm{ \mathcal Q}_i,
\end{equation}
where $\bm S_i$ is the electron spin at the site $i$, $\bm \Omega_i = \bm \Omega_{i,{\rm nucl}} + \bm \Omega_B$ is the electron spin precession frequency caused by the hyperfine interaction with nuclear spins, $\bm \Omega_{i,{\rm nucl}}$, and by the Larmor precession in the external field, $\bm \Omega_B$. The collision integral $\bm{ \mathcal Q}_i$ describes the variations of the spins due to the electron hopping between sites, processes of ionization and recombination, exchange diffusion, etc.~\cite{toddkarin:opticalOrientationBook,toddkarin:Kavokin2008}. The schematic illustration of the spin dynamics of localized electrons is presented in Fig.~\ref{fig:mechanism}(a). Here we employ the simplest model of the collision integral by introducing a single correlation time $\tau_c$, disregarding the spread of the transition probabilities~\cite{toddkarin:opticalOrientationBook,PhysRevB.91.195301}. We assume  that the nuclear fluctuations are frozen on the timescale of $\tau_c$ and that the Zeeman splitting in the external field is negligible as compared with the thermal energy. Hence, we obtain a simple analytical formula for the relaxation time of the spin component parallel to the magnetic field ${\mathbf B\parallel z}$~\cite{PhysRevB.91.195301}:
\begin{equation}
\label{T1:nucl}
T_{1,hf}  = \frac{\tau_c \mathcal A}{1-\mathcal A},
\end{equation}
where 
\begin{equation}
\label{A:rel}
\mathcal A = \left\langle \frac{1+ \Omega_{i,z}^2\tau_c^2}{1+\Omega_i^2\tau_c^2} \right\rangle,
\end{equation}
and the angular brackets denote the averaging over the distribution of random nuclear fields.

Equation~\eqref{T1:nucl} is valid for an arbitrary relationship between the spin precession frequency and $\tau_c$. In the experimentally relevant range of magnetic field, ${\Omega_B = |g\mu B|/\hbar}$ exceeds by far the spin precession frequency in the field of nuclear fluctuations and the inverse correlation time. It follows then from Eqs.~\eqref{T1:nucl}, \eqref{A:rel} that
\begin{equation}
\label{B2}
T_{1,hf}=  \frac{3\tau_c\Omega_B^2}{2\langle \Omega_{\rm nucl}^2\rangle}\propto \tau_c B^2,
\end{equation}
where $\langle \Omega_{\rm nucl}^2\rangle$ is the mean square fluctuation of the nuclear field averaged over the ensemble of donors.
This expression shows the $B^2$ power law which is observed in experiment, Fig.~\ref{fig:T1expdata}. 

This increase in spin-relaxation time with increasing field is related to the suppression of the relaxation by the magnetic field: At ${\Omega_B \gg \tau_c^{-1},\langle \Omega_{\rm nucl}^2\rangle^{1/2}}$, the electron spin precesses around the total field ${\bm \Omega_{B} + \bm \Omega_{i,{\rm nucl}}}$ during the correlation time. Its precession axis is almost parallel to $\bm \Omega_B$ and its orientation changes by a small random angle $\sim \Omega_{i,{\rm nucl}}/\Omega_B$ when the electron hops between the localization sites. Such a random process results in the spin relaxation rate $\sim \tau_c^{-1}(\Omega_{i,{\rm nucl}}/\Omega_B)^2 \propto 1/(\tau_cB^{2})$ in agreement with Eq.~\eqref{B2}.
For known mechanisms of electron correlation time at a donor, such as electron hopping and the exchange diffusion, see Ref.~\cite{toddkarin:Kavokin2008} for review, an exponential sensitivity to the donor density (and, in the former case, to the temperature) is expected~\cite{toddkarin:Kavokin2008,ES:book}. Correspondingly, for these mechanisms $T_1$ should be strongly affected by these parameters. Such trends are clearly seen in the experiment, Fig.~\ref{fig:T1expdata} and Fig.~\ref{fig:tempdataInP}(b).

The developed model enables quantitative comparison with the experiment. To that end, we evaluate the mean square of the donor-bound electron spin precession frequency in the nuclear field as~\cite{dp74} 
\begin{equation}
\label{donor}
\langle \Omega_{\rm nucl}^2 \rangle= \frac{{V}_0}{8\pi (a_B^*)^3 \hbar^2}\sum_\alpha {\left(A_\alpha^{hf}\right)^2} I_\alpha(I_\alpha+1),
\end{equation}
where ${a_B^*{=\varepsilon \hbar^2/(m^*e^2)}}$ is the donor Bohr radius, ${V_0 = a_0^3}$ is the unit lattice volume, $I_\alpha$ is the spin of $\alpha^{\rm th}$ nucleus in a unit cell, $A_\alpha$ is the hyperfine interaction constant.
Taking for GaAs ${A_{^{69}\rm Ga} = 38.2~\mu\text{eV}}$, ${A_{^{71}\rm Ga} = 48.5~\mu\text{eV}}$ and ${A_{^{75}\rm As} = 46~\mu\text{eV}}$~\cite{toddkarin:Syperek2011} we obtain $\sqrt{\langle \Omega_{\rm nucl}^2 \rangle} = 0.47\times 10^8~\mbox{s}^{-1}$. Fitting the experimental data with Eq.~\eqref{B2}, we determine a correlation time $\tau_c \approx 25$~ns for the GaAs-2 sample. Such a value of the correlation time is consistent with previous studies of GaAs samples with similar donor densities~\cite{toddkarin:Dzhioev2002,toddkarin:Berski2015}. A somewhat longer $\tau_c$ of $\sim 40$~ns is determined for the InP-2 sample, where the hyperfine interaction is dominated by $^{115}$In isotopes with ${I_{\rm In} = 9/2}$. The estimate for $A_{\rm In}$ comes from Ref.~\cite{PhysRev.135.A200} where the Overhauser effect for InSb was measured. The literature reports a spread of $A_{\rm In}$: $ 47$~$\mu$eV~\cite{toddkarin:Chekhovich2011}, $56$~$\mu$eV~\cite{PhysRevB.79.195440} and $84$~$\mu$eV~\cite{toddkarin:Syperek2011}. Here we use the middle value of ${A_{\rm In} = 56~\mu\text{eV}}$, which yields $\sqrt{\langle \Omega_{\rm nucl}^2 \rangle} = 1.6\times 10^9~\mbox{s}^{-1}$.

Although the experimental sensitivity of \Tone to temperature and carrier density are consistent with the known mechanisms contributing to the donor electron correlation time, the magnitude of $\tau_c$ is orders of magnitude shorter than these mechanisms predict for the low donor densities used in this study. Our result is consistent with prior works~\cite{toddkarin:Dzhioev2002,toddkarin:Berski2015} and suggests additional, unknown mechanisms may be at play, such as an inhomogeneous donor distribution resulting in the formation of clusters with a relatively high donor density, and short $\tau_c$.

According to Eq.~\eqref{B2}, the electron spin relaxation time associated with the hyperfine interaction strongly increases with an increase in field. Hence, at sufficiently strong magnetic fields this mechanism becomes inefficient as compared with mechanisms caused by the combination of the electron-phonon and spin-orbit interactions described below. By contrast, \Tone due to these processes decreases with an increase in $B$.

\subsection{High-field spin-relaxation}\label{sec:highFieldAdmixture}

While the spin-orbit interaction alone is not sufficient to cause a spin-flip of a localized charge carrier, a combination of the electron-phonon interaction and spin-orbit coupling serves as a main source of localized electron spin relaxation at high magnetic fields~\cite{toddkarin:Pines1957,toddkarin:Abrahams1957,toddkarin:Frenkel1991}. Phonons can also modulate the hyperfine coupling of the electron and the lattice-nuclei spins giving rise to ${T_1\propto B^{-3}}$ dependence~\cite{toddkarin:Pines1957}. Similar to the quantum dot case, this effect is negligible for donor-bound electrons. Two-phonon processes~\cite{toddkarin:Abrahams1957} are also very weak for the range of temperatures and fields studied here.

An exhaustive theoretical investigation of the spin-flip mechanisms has been performed for the related GaAs quantum dot system~\cite{toddkarin:Khaetskii2001,PhysRevB.66.161318,toddkarin:Marquardt2005}. 
In GaAs quantum dots, all reported spin-orbit related mechanisms exhibit a ${T_1 \propto B^{-\nu}}$ dependence with $\nu\geq5$. For bulk GaAs-like semiconductors, such a study has not been performed before to the best of our knowledge.
The orbitals for the donor-bound electron differ from those for quantum dots, leading to the use of a different approximation for the Dresselhaus spin-orbit Hamiltonian and different selection rules.

Experimentally we observe that the high-field spin relaxation is consistent with a single phonon process. This limits us to mechanisms that combine Dresselhaus spin-orbit coupling and {spin-conserving phonon-induced relaxation, and direct spin-phonon mechanisms.  In this section, we present the detailed calculation for the high-field \Tone due to both mechanisms and compare our theoretical results to the experimental data.

\subsubsection{Admixture mechanism caused by Dresselhaus spin-orbit coupling}

\begin{table}[]
 \caption{Material parameters relevant to the donor-bound electron spin-relaxation in GaAs, InP, and CdTe. \(g\) is the effective electron \(g\)-factor, \(m^*\) is the electron effective mass ($m_0$ is the free electron mass), \(h_{14}\) is the piezoelectric constant, \(\gamma\) is the Dresselhaus spin-orbit coupling constant, \(\rho\) is the mass-density, \(s_l\) is the longitudinal sound velocity, \(s_t\) is the transverse sound velocity, \(\varepsilon\) is the relative permittivity of the material, \(v_0\) characterizes the strength of the direct spin-phonon coupling interaction, and $D$ is the deformation potential interaction constant.}
 \label{table:parameters}
    \centering
 \begin{tabular}{cccc}
 \hline
 \hline
            & GaAs & InP  & CdTe \\
      \hline
 $g$ & $-0.44$ & 1.3 & $-1.67$ \\
 $m^*$ & $0.067 m_0$ & $0.08 \, m_0$ &  $0.106 m_0$ \cite{toddkarin:Petrovic2012} \\
 $h_{14}$ (V/m) & $14.5 \times 10^8$  \cite{toddkarin:Madelung2004} & $7.4 \times 10^8$ \cite{xiayu:Rottner1993} & $3.94 \times 10^8$ \cite{toddkarin:Petrovic2012} \\
 $\gamma$ (eV$\cdot${\AA}$^3$) & 23.7 \cite{toddkarin:Jancu2005} & 10.1 \cite{toddkarin:Jancu2005} & 11.74 \cite{toddkarin:Petrovic2012} \\
 $\rho$ (kg/m$^3$) & $ 5.32 \times 10^3$ \cite{toddkarin:ioffeWebsite} & $ 4.81 \times 10^3 $ \cite{toddkarin:ioffeWebsite} & $4.85 \times 10^3 $ \cite{toddkarin:Petrovic2012} \\
 $s_{l}$ (m/s) & $ 4.73 \times 10^3$   \cite{toddkarin:ioffeWebsite} & $ 4.58 \times 10^3 $ \cite{toddkarin:ioffeWebsite} & $3.08 \times 10^3  $ \cite{toddkarin:Petrovic2012} \\
 $s_{t}$ (m/s)  & $ 3.35 \times 10^3 $ \cite{toddkarin:ioffeWebsite}  & $ 3.08 \times 10^3 $ \cite{toddkarin:ioffeWebsite}  & $1.85 \times 10^3 $ \cite{toddkarin:Petrovic2012} \\
{$\varepsilon$} & $12.56$ & $12.5$ \cite{toddkarin:ioffeWebsite} & $\sim10.3$~\cite{lbIIVI}\\
$v_0$ (m/s) & 8$\times10^5$ \cite{toddkarin:Dyakonov1986} & $4\times10^5$ \cite{toddkarin:Dyakonov1986} & unknown \\
 $D$ (eV) & $-5.55$ \cite{toddkarin:Gavini1970} & $-4.4$ \cite{toddkarin:Gavini1970}  &  $-5.45$ \cite{toddkarin:Gavini1970} \\
\hline
\hline
 \end{tabular}
\end{table}

\begin{figure*}[ht]
\includegraphics[width=\linewidth]{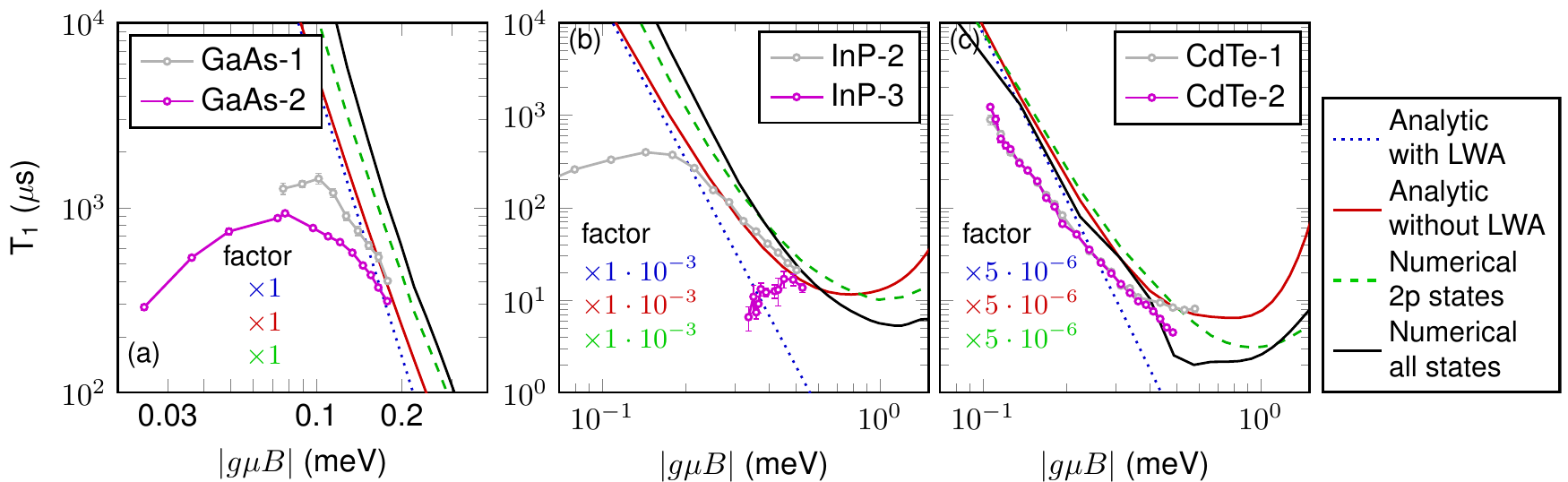}
\caption{Theoretical results for spin-relaxation time \Tone via the admixture mechanism, using both analytic and numerical wave functions. Pink and grey dots show the experimental data. For GaAs [panel (a)], the theory matches the data reasonably well with no fit parameters. For InP (b) and CdTe (c), the calculated values are multiplied by the factor specified in the figure for ease of comparison. ${T=1.5~\text{K}}$. We note that the numerically calculated \Tone which includes only the 2p states is slightly shorter than the full numerical solution. This is due to destructive interference between the orbital states in Eq.~\eqref{eq:M}. 
}
\label{fig:T1theorydata}
\end{figure*}

We are first interested in the spin relaxation between the Zeeman sublevels of the donor-bound electron ground state mediated by spin-orbit and electron-phonon coupling (admixture mechanism). This is the dominant relaxation mechanism for III-V quantum dots~\cite{toddkarin:Khaetskii2001,PhysRevB.66.161318} and naively may also be expected to play the dominant role in the similar donor system. For this mechanism, the spin-orbit interaction modifies the ground-state Zeeman sublevels by the admixture of the excited sublevels with the opposite spin component. Hence, the spin-independent electron-phonon coupling causes spin-relaxation through the components of the states with the same spin, as depicted in Fig.~\ref{fig:mechanism}(b)-(c).

The interaction Hamiltonian for the admixture mechanism is
\begin{equation}
\label{H:adm}
    H_{adm} = U_{ph} + H_{so},
\end{equation}
where $U_{ph}$ is the spin-conserving  electron-phonon interaction Hamiltonian and $H_{so}$ is the spin-orbit Hamiltonian. In the high-field limit, the Zeeman splitting can be comparable or even exceed the thermal energy. In such a case, the transition rates from the Zeeman sublevel~$\downarrow$ to~$\uparrow$, \(\Gamma_{\uparrow \downarrow}\), and back, \(\Gamma_{\downarrow \uparrow}\), differ. The observed longitudinal spin relaxation time satisfies 
\[{T_1 = (\Gamma_{\uparrow \downarrow} + \Gamma_{\downarrow \uparrow } )^{-1} }.\]
The individual rates are found using Fermi's golden rule, e.g.,
\begin{equation}\label{eq:T1start}
\Gamma_{\downarrow\uparrow} = \frac{2 \pi}{\hbar} \sum_{\mathbf{q},\alpha} |M_{\downarrow\uparrow}|^2 \delta(\hbar q s_\alpha - |g \mu B|),
\end{equation}
where $\mathbf q$ is the phonon wavevector, \(s_\alpha\) is the speed of sound in phonon branch \(\alpha\) and $\alpha=t,l$ for the transverse and longitudinal modes, respectively. Hereafter we assume for convenience that the spin-up state has higher energy than the spin-down one, hence, ${g\mu B>0}$, as illustrated in Fig.~\ref{fig:mechanism}, so that the rate in Eq.~\eqref{eq:T1start} corresponds to the phonon emission process. Electron spin-relaxation occurs via a second order process due to the quantum interference of $U_{ph}$ and $H_{so}$ in the Hamiltonian~\eqref{H:adm}, see Ref.~\cite{toddkarin:Khaetskii2001} for details,
\begin{eqnarray}\label{eq:M}
\nonumber M_{\downarrow\uparrow, adm} = -\sum_{e} \bigg[ \frac{\langle \text{1s}, \downarrow |U_{ph}|e, \downarrow \rangle \langle e, \downarrow |H_{so}|\text{1s}, \uparrow \rangle}{E_e - E_{\text{1s}} +  g \mu B} \\
+\frac{\langle \text{1s}, \downarrow |H_{so}|e,  \uparrow \rangle \langle e,  \uparrow |U_{ph}|\text{1s}, \uparrow \rangle }{E_e - E_{\text{1s}} -  g \mu B}  \bigg],
\end{eqnarray}
where $|\text{1s}\rangle$ is the ground orbital state of the donor-bound electron, $|e\rangle$ denotes the excited orbital states, and $E_e$, $E_{\text{1s}}$ are the energies of the corresponding orbitals.

Due to the small localization energy of the donor-bound electron ($\lesssim10$~meV), the electron wave function in a magnetic field is well described with effective mass theory using the hydrogenic Hamiltonian
\begin{equation}
H_{0} = \frac{\hbar^2}{2m^*} \left(\mathbf{k} - \frac{e}{\hbar}\mathbf{A}\right)^2  -  \frac{1}{4 \pi  \varepsilon_0 }\frac{e^2}{ \varepsilon r} + \frac{1}{2} g \mu  \bm \sigma \cdot {\mathbf B},
\label{eq:Ho}
\end{equation}
where $m^*$ is the electron effective mass, $e$ is the electron charge, $\mathbf{A}$ is the vector potential of the magnetic field $\mathbf B$, $\mathbf r$ is the position vector, ${r=|\mathbf r|}$, ${\mathbf k = - i \partial/\partial \mathbf r}$ is the wavevector, \(\varepsilon\) is the relative dielectric constant of the material, and ${\bm \sigma}$ is the vector composed of the Pauli matrices. 
In the presence of the magnetic field, the Hamiltonian, Eq.~\eqref{eq:Ho}, possesses an axial symmetry and its eigenstates are characterized by four quantum numbers: principal quantum number \(\nu\), angular momentum $z$-projection \(m\),  $z$-parity \(\pi_z\) and spin $z$-projection \(m_s\). To establish a link with the hydrogen-like series of donor-bound electron states at $B=0$, we will label the orbitals by their zero-field quantum numbers $nlm$, where $n$ is the principal quantum number and $l$ is the angular momentum quantum number, when appropriate.

The energy of a phonon involved in the spin-flip transition is the Zeeman splitting between the spin sublevels. Therefore, the phonon wavevector ${q_{\alpha}=g\mu B/(\hbar s_{\alpha})\to 0}$ as ${B\to 0}$. Thus, at moderate magnetic fields in piezoelectric crystals such as GaAs, InP and CdTe studied here, we found that the piezoelectric electron-phonon interaction with ${U_{ph}^{(pz)} \propto q^{-1/2}}$ dominates over the deformation potential interaction, where ${U_{ph}^{(dp)} \propto q^{1/2}}$~\cite{toddkarin:Gantmakher1987}, see Appendix~\ref{appendix:admixtureTheory}.  The piezoelectric electron-phonon interaction reads
\begin{equation}
\label{U:piezo}
U_{ph}^{(pz)} = \sqrt{\frac{\hbar}{2\rho \omega_{\mathbf{q},\alpha}}}e^{i (\mathbf{q r} - \omega_{\mathbf q,\alpha} t)} (e A_{\mathbf{q},\alpha}) b^{\dagger}_{\mathbf{q},\alpha} + {\rm c.c.},
\end{equation}
where
\begin{equation}
A_{\mathbf{q}, \alpha} = h_{14}\sum_{ijk} \beta_{ijk}\xi_i \xi_j \hat{e}_k^{(\mathbf{q},\alpha)},
\end{equation}
$\rho$ is the mass density of the material, $\omega_{\mathbf q, \alpha}$ is the phonon frequency,  $b^\dagger_{\mathbf q,\alpha}$ is the creation operator for a phonon, $\bm \xi = \mathbf q/q$ is the unit vector along the phonon wavevector, $\hat{\bm e}$ is the phonon polarization vector, the only nonzero components of $\beta_{ijk}$ are those with different subscripts, $\beta_{xyz} =  \ldots = \beta_{zyx}=1$, and $h_{14}$ is the piezoelectric constant~\cite{toddkarin:Gantmakher1987}.

Since all the samples studied here are bulk semiconductors characterized by the $T_d$ point symmetry group, the only relevant spin-orbit coupling comes from the cubic-in-the-electron-wavevector Dresselhaus spin-orbit term, $H_{so}$. 
It arises from the lack of inversion symmetry in zinc-blende crystals and has the form
\begin{equation}\label{eq:Hso}
H_{so} = \gamma \sum_i \sigma_i k_i(k_{i+1}^2-k_{i+2}^2),
\end{equation} 
where \(\gamma\) is the Dresselhaus spin-orbit coupling constant and the subscript $i$ cycles through $x$, $y$, $z$. 

Depending on the relation between the magnetic length, ${l_b=\sqrt{\hbar/|eB|}}$, and the effective Bohr radius, ${a_B^*}$, various regimes of the spin-flip can be realized. At sufficiently weak magnetic fields, where ${l_b \gg a_B^*}$, the magnetic field does not affect the hydrogen-like states of the donor-bound electron. In this case, the Dresselhaus spin-orbit interaction admixes $n$f-shell states with principal quantum numbers ${n=4,5, \ldots}$ and orbital momentum ${l=3}$ to the 1s-shell state. With the long wavelength approximation (LWA) for the phonons, where ${|g \mu B| a_B^*/(\hbar s_\alpha) \ll 1}$, we obtain the longitudinal spin relaxation time 
\begin{equation}
\label{lowlowB9}
T_{1,adm}^{(low)} \propto B^{-9} F^{-1}_{ph},
\end{equation}
see Appendix~\ref{appendix:admixtureTheory} for details. Hence, at low temperatures, $T_1$ is inversely proportional to $B^9$, while for $k_B T \gtrsim g\mu B$, $T_1 \propto B^{-8}$. We do not observe this regime in experiments due to the dominating low-field nuclear-electron hyperfine mechanism.

In the opposite limit, where ${l_b \ll a_B^*}$ and, moreover, ${\hbar\omega_c \gg \mathcal E_{Ry}^*}$, where ${\omega_c = |eB/m^*|}$ is the cyclotron frequency and ${\mathcal E_{Ry}^*= {m^*e^4/[2(4\pi \varepsilon\varepsilon_0)^2\hbar^2]}}$ 
is the donor-bound-electron binding energy, the magnetic field shrinks the wave functions of the ground and excited states of the donor-bound electron. This situation is similar to the case of an electron localized in the $(xy)$ plane by a parabolic potential, like in the quantum dot system studied in Ref.~\cite{toddkarin:Khaetskii2001}. Here the excited states with $|m|=1$ (in addition to those with $|m|=3$) are admixed and, in the LWA, we obtain for the spin flip time
\begin{equation}
\label{highhighB3}
T_{1, adm}^{(high)} \propto B^{-3} F^{-1}_{ph},
\end{equation}
see Appendix~\ref{appendix:admixtureTheory}. 
This high-field limit is not realized for the studied samples and magnetic fields accessible in our experiments. Moreover, in this limit, the LWA in our system is no longer valid.

%
%
%
%
Therefore, we have performed the full numerical evaluation of the spin relaxation time according to Eqs.~\eqref{eq:T1start} and \eqref{eq:M} using the numerical solutions to Eq.~\eqref{eq:Ho}~\cite{toddkarin:Schimeczek2014} and the material parameters from Table~\ref{table:parameters}. Additional details on the numerical calculation can be found in Appendix~\ref{appendix:h2db}. These results, which include 18 excited state orbitals, are given by the black curves in Fig.~\ref{fig:T1theorydata}. We numerically find that the first excited state which evolves from 2p$_-$ makes the dominant contribution to the spin relaxation rate, as shown by the dashed green curves in Fig.~\ref{fig:T1theorydata}.

This numerical result, together with the analysis of the wavefunctions in Appendix~\ref{appendix:admixtureTheory}, motivates using Gaussian shapes of the ground and excited state wave functions, Eqs.~\eqref{eq:app_wfs}, to obtain an analytic solution for further insight into the intermediate field behavior.
The magnetic field induced shrinking is taken into account by assuming different characteristic lengths ${l_{z,\text{1s}} = a_B^*}$ and ${l_{\rho,\text{1s}}=[1/(a_B^*)^2+1/(2l_b^2)]^{-1/2}}$ for the motion along and perpendicular to the field. After some transformations, we obtain
(see Appendix~\ref{appendix:admixtureTheory} for details):
\begin{multline}
\label{Tone:adm:analyt}
\frac{1}{T_{1, adm}} = {\frac{256 \chi^{10}}{35(1+\chi^2)^{12}}}\frac{\gamma^2 e^4 h_{14}^2 {|}g \mu{|}^3 B^5}{ \pi \rho \hbar^6} \times \\
\times
\left(\frac{1}{\Delta E}-\frac{1}{\Delta E + \hbar \omega_c}\right)^2 
\left( \frac{f_l}{s_l^5}  + \frac{4f_t}{3s_t^5}  \right) F_{ph}\:.
\end{multline}
Here, $\Delta E = E_{\text{2p}_-} - E_{\text{1s}}$ is the energy difference between the hydrogen-like ground 1s and excited $\text{2p}_-$ state.
The factors $f_{\alpha} = \exp{\{-(\chi g\mu B{l_\rho})^2/[(1+\chi^2)\hbar^2s_{\alpha}^2]\}}$ take into account that the phonon wavelength can be comparable with the donor-bound electron state size. These factors are particularly sensitive to the wavefunction shape. Finally, the parameter $\chi$ is a parameter of the wave functions which characterizes the ratio of the effective radii for the excited and the ground states, see Eqs.~\eqref{eq:app_wfs}. By comparing the trial wavefunction to the numerical 2p wavefunction, we find reasonable choices for $\chi$ of 1.5, 1.7 and 2.2 for GaAs, InP and CdTe over the experimental range of magnetic field, as shown in Fig.~\ref{fig:numericChi}. The magnitude of $T_{1,adm}$ calculated according to Eq.~\eqref{Tone:adm:analyt} is quite sensitive to the choice of $\chi$. 

A comparison between the experimental, numerical, and analytic results for \Tone is shown in Fig.~\ref{fig:T1theorydata}. We stress that these calculations contain no fitting parameters. Qualitatively we observe similar behavior between the analytic and numerical calculations. At sufficiently strong magnetic fields, we find the LWA fails for InP and CdTe due to their relatively large  electron $g$-factors as compared with GaAs. This effect is taken into account by factors $f_l$ and $f_t$ in Eq.~\eqref{Tone:adm:analyt}. It softens the exponent in $B$-dependence giving approximately $3 \lesssim \nu \lesssim 4$ in the accessible field range. Further increase in $B$ results in a minimum in $T_1(B)$. It is noteworthy that at such magnetic fields, the deformation potential interaction may become important, see~\cite{PhysRevB.72.125340} and Appendix~\ref{appendix:admixtureTheory} for details; moreover, in such fields the result could be quite sensitive to the shape of the wave functions. Hence, for sufficiently high fields, Eq.~\eqref{Tone:adm:analyt} provides only an indication of the trend.

We find that the numerically calculated values of \Tone for InP and CdTe are orders of magnitude longer than the experimentally observed spin relaxation times in these samples, which indicates the importance of other spin-flip mechanisms in the materials, see below. By contrast, in GaAs the calculated magnitude of \Tone is quite close to the experimental values, demonstrating that the admixture mechanism is significant in this material.



\subsubsection{Direct spin-phonon interaction}\label{sec:direct}

\begin{figure*}[ht]
    \centering
    \includegraphics[width=\linewidth]{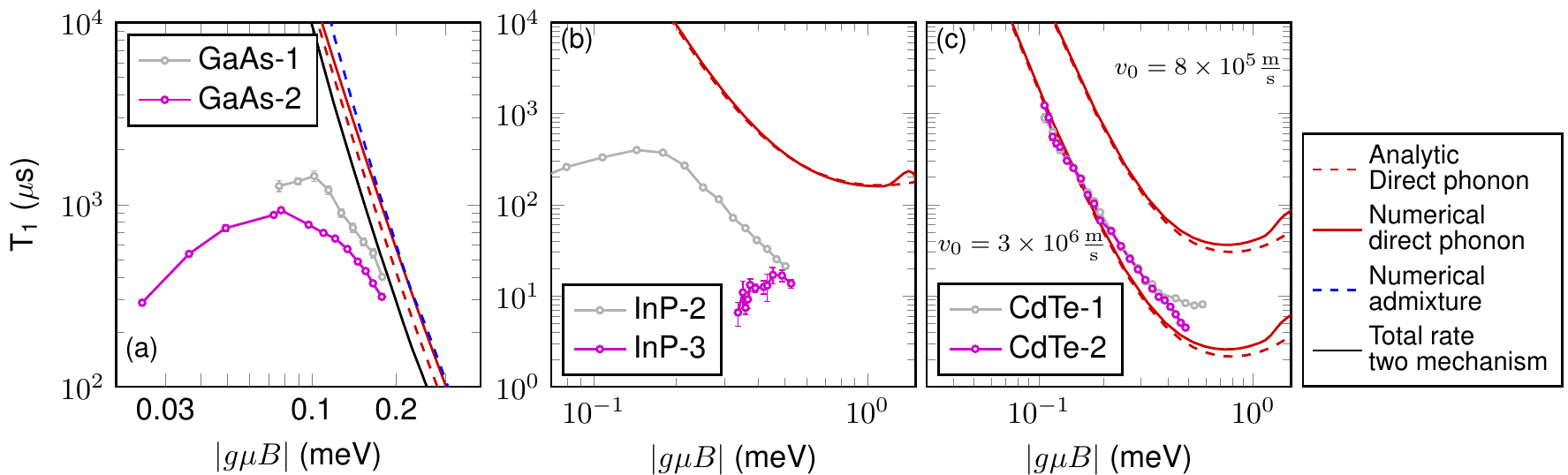}
    \caption{Theoretical results for the spin-relaxation time \Tone via the direct spin-phonon mechanism for GaAs~(a), InP~(b) and CdTe~(c). Pink and grey dots show the experimental data. $T=1.5~\text{K}$. The two dashed lines and two solid lines in (c) represent the analytic and numerical calculation results of $T_1$ using ${v_0=8\times10^5~\text{m/s}}$ (upper curves) and ${v_0=3\times10^6~\text{m/s}}$ (lower curves)}
    \label{fig:directphonon}
\end{figure*}

Although the direct spin-phonon interaction was not found to be a dominant relaxation mechanism for electrons in semiconductor quantum dots~\cite{toddkarin:Khaetskii2001}, we demonstrate here that it contributes significantly to donor-bound electron spin relaxation. To some extent, this is because the role of the admixture mechanism is diminished due to the cubic-in-the-wavevector spin-orbit splitting in the bulk material, as compared with $\mathbf k$-linear terms used for quantum dot systems~\cite{toddkarin:Khaetskii2001}. The direct spin-phonon interaction Hamiltonian is~\cite{toddkarin:Dyakonov1986} 
\begin{multline}
   U_{dir} = \frac{\hbar v_0}{2}  [\sigma_x (u_{xy}k_y - u_{xz}k_z) + \\
   + \sigma_y (u_{yz} k_z - u_{yx} k_x) + \sigma_z (u_{zx} k_x - u_{zy} k_y)],
    \label{eq:Uph:0}
\end{multline}
Here $u_{ij}=u_{ji}$ is the deformation tensor, and, as above, ${\mathbf k = -i \bm \nabla - (e/{\hbar}) \mathbf A}$. The coupling constant $v_0$  has the dimension of velocity. It has been determined by experiment for GaAs and InP but is unknown for CdTe (see Table~\ref{table:parameters}). For numerical evaluation for CdTe, we use a spread of values, ${8\times10^5~\mathrm{m/s}}<{ v_0^{\rm CdTe}}< {3\times10^6~\mathrm{m/s}}$ with the lower (upper) bound corresponding to $v_0^{\rm GaAs}$ $\left(v_0^{\rm InSb}\right)$~\cite{toddkarin:Dyakonov1986}. 

The relaxation rates $\Gamma_{\uparrow\downarrow}$ are calculated using Eq.~\eqref{eq:T1start} with the first-order matrix element ${M_{\uparrow
\downarrow}= \langle \text{1s},\uparrow | U_{dir} | \text{1s},\downarrow\rangle}$, as depicted in Fig.~\ref{fig:mechanism}(c). We use an approximate exponential wave function with a characteristic length ${l = [{(a_B^*)}^{-2} + 1/(2l_b^2)]^{-1/2}}$ to obtain analytic expressions for $\Gamma_{\uparrow\downarrow}$ and, correspondingly, for the associated longitudinal spin relaxation time $T_{1,dir}$. The choice of the wave function is motivated by the fact that only the $\text{1s}$ orbital state is involved, which is not significantly perturbed at the experimentally accessible magnetic fields. Moreover, the precise symmetry of the wave function for the direct phonon mechanism is not critical. The evaluation of Eq.~\eqref{eq:T1start} yields (see Appendix~\ref{appendix:direct}):
\begin{equation}
    \begin{aligned}
        & \frac{1}{T_{1,dir}}   =   \frac{(e v_0 l^2)^2 |g \mu|^5 |B|^7}{560 \pi\rho \hbar^6}F_{ph} \times \\
    & \left( \frac{1}{s_l^7}\frac{1}{(1+Q_l^2)^6}+\frac{4}{3 s_t^7}\frac{1}{(1+Q_t^2)^6} \right)  ,
      \end{aligned}
      \label{eq:direct}
\end{equation}
where ${Q_{\alpha}= |g\mu B| l/(2\hbar s_{{\alpha}})}$. 
Equation~\eqref{eq:direct} demonstrates that the spin-flip time is proportional to $B^{-7}$ at weak magnetic fields. An increase in the field results in a softening of the $B$-field dependence due to decrease of the efficiency of the electron-phonon interaction (breakdown of the LWA) described by the factors $(1+Q_\alpha^2)^{-6}$.
In addition to the analytic approximation, we performed the full calculation using the numerically-obtained ground-state donor wave function. The very good agreement between the analytic and numerical calculations, seen in Fig.~\ref{fig:directphonon}, can be attributed to the minor effect of the magnetic field on the ground-orbital-state wave functions at the experimental fields.  

A comparison between the theoretical calculations with no fit parameters and the experimental data is also provided in Fig.~\ref{fig:directphonon}. For GaAs, we find that the magnitude of the direct-phonon mechanism is approximately the same as the admixture mechanism. Also included in Fig.~\ref{fig:directphonon}(a) is the sum of these two mechanisms. Accounting for both mechanisms results in a difference between the theory and the data of approximately a factor of 2, which can be easily attributed to the uncertainties in the system parameters in Table~\ref{table:parameters}. 

For InP and CdTe, the direct spin-phonon mechanism is found to be significantly stronger than the admixture mechanism. For CdTe, the agreement between theory and experiment is extremely good if the direct spin-phonon interaction strength in CdTe is similar to that of InSb. This may be reasonable given the similar valence band spin-orbit splitting in the two materials, 0.8~eV in InSb~\cite{ref:cardona1988rbs} and 0.9~eV in CdTe~\cite{ref:twardowski1980vbs, ref:marple1962dcb,ref:cardona1967ese}. Here, an independent measurement of $v_0$, like those performed in Ref.~\cite{toddkarin:Dyakonov1986} for GaAs and InP, or its independent first-principles calculation, is needed to corroborate our result.

There is still a significant discrepancy between theory and experiment for InP, where the experimental spin-relaxation time is 15 to 30 times shorter than the predicted value from the direct spin-phonon coupling. Its origin is not clear and further studies, both experimental and theoretical, are needed to resolve this discrepancy.

\section{Conclusion}\label{sec:conclusion}

In this work we measure the longitudinal spin relaxation time as a function of magnetic field for electrons bound to donors in three different high-purity direct bandgap semiconductors. We observe for the first time the crossover
between low-field spin relaxation resulting from a hyperfine coupling of the electron and lattice nuclear spins and high-field single-phonon-mediated spin relaxation. From a fundamental perspective, the existence of both regimes is expected. However, the comparison of the data with the developed theory in terms of the magnitude of the relaxation raises new questions. Low field measurements indicate a tens of nanoseconds electron spin correlation time of so far unknown origin. High-field measurements strongly suggest the admixture mechanism is important in GaAs, while the direct spin-phonon interaction is important in both CdTe and GaAs. However for InP, the discrepancy between theory and experiment calls for further investigation.

In the context of possible applications, the high-field $B^{-\nu}$ dependence of \Tone, combined with the density and temperature dependent low-field $B^{2}$ behavior, has practical implications. If the 
crossover point can be pushed to lower fields, extremely-long spin-relaxation times may be possible. This could be realized with lower impurity density, lower temperature, larger binding energies, and a nuclear-spin-free matrix. In support of this, we note that no crossover is observed in CdTe even when ${k_BT >|g\mu B|}$. This may reflect the role of the higher donor binding energy and/or the reduced nuclear-spin environment in CdTe. In this context, isotope purification, which is known to significantly affect spin dephasing, may also significantly increase the maximum achievable $T_1$ for electrons bound to shallow donors. 

\section{Acknowledgements}
The authors thank P. Rivera, P. Wilhelm, and B. Ebinger for assistance with building the measurement apparatus and E.L. Ivchenko for valuable discussions. We thank Colin Stanley for the MBE GaAs samples provided via the University of Glasgow. This material is based upon work supported by the National Science Foundation under Grant Number 1150647 and the National Science Foundation Graduate Research Fellowship under grant number DGE-1256082. E.Y.S. acknowledges support of the University of the Basque Country UPV/EHU under program UFI 11/55, Spanish MEC (FIS2012-36673-C03-01 and
FIS2015-67161-P) and Grupos Consolidados UPV/EHU del Gobierno Vasco
(IT-472-10). M.M.G. was partially supported by RFBR project No.~14-02-00168, the Russian Federation President Grant MD-5726.2015.2, Dynasty foundation. M.V.D. was partially supported by RFBR project No.~16-32-60175 and the Dynasty foundation.



\appendix
\setcounter{figure}{0}
\renewcommand{\thefigure}{\Alph{section}\arabic{figure}}

\section{Magneto-photoluminescence spectra for GaAs, InP, and CdTe}
\label{appendix:magnetoPL}
Representative magneto-photoluminescence spectra for GaAs-2, InP-2, and CdTe-2 are shown in Fig.~\ref{fig:magnetoPL}. In all three samples we can observe the free exciton (labeled X), donor-bound exciton $\text{D}^0\text{X} \to \text{D}^0\text{,1s}$ transition (labeled D$^0$X), ionized donor-bound exciton transition $\text{D}^+\text{X} \to \text{D}^+$ (labeled D$^+$X), and acceptor-bound exciton $\text{A}^0\text{X} \to \text{A}^0\text{,1s}$ transition (labeled A$^0$X). Also observed in GaAs and InP are the D$^0$X two-electron satellite (TES) transitions which correspond to the
$\text{D}^0\text{X} \to \text{D}^0,nl^{m}$ transition, where \(n,l,m\) specify the quantum numbers of the excited D$^0$ orbital at $B=0$.
For GaAs and InP, the fine-structure of the D$^0$X spectra is well resolved due to the hole spin and spin-orbit interaction as well as the nearby D$^0$X excited orbital states. In the CdTe samples, which are bulk crystals, this structure is unresolved, limiting our ability to optically pump the system to electron Zeeman splittings greater than 0.1~meV.

\begin{figure}
    \includegraphics{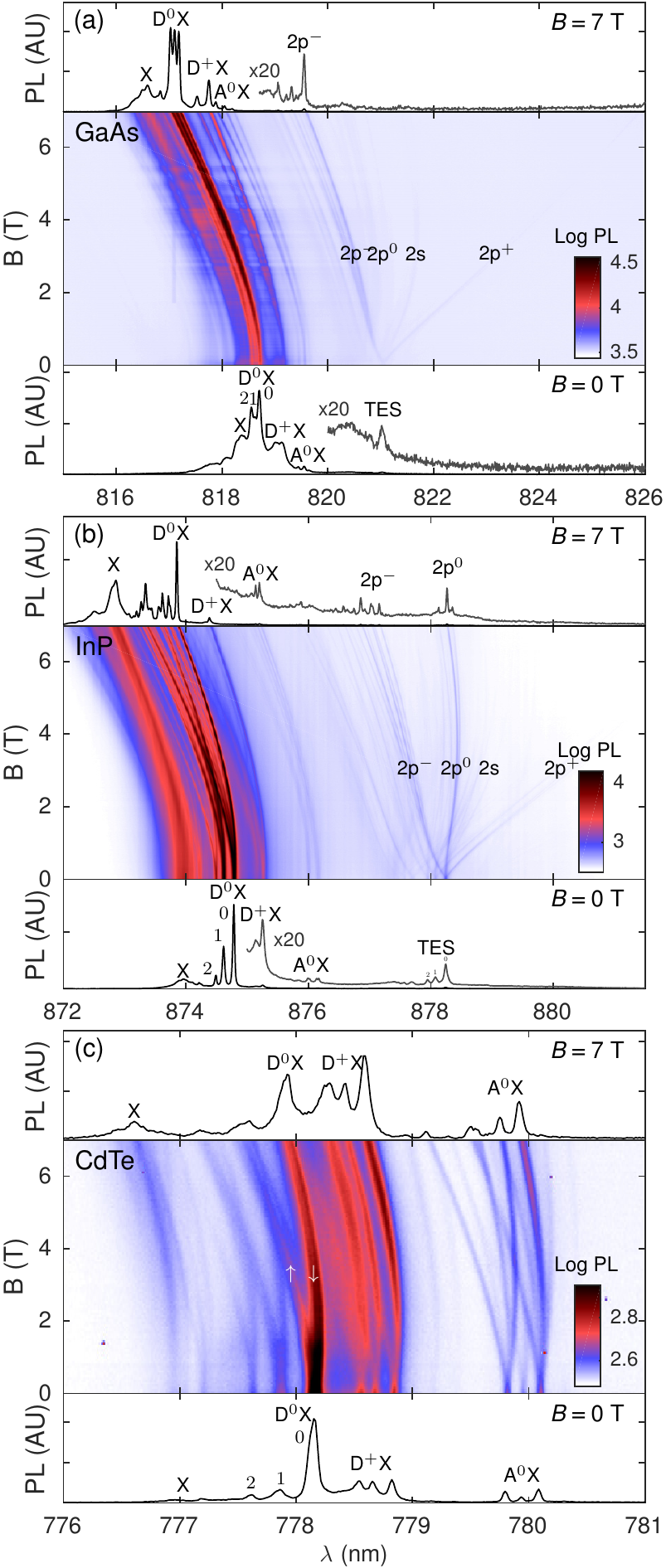}
    \caption{Magneto-photoluminescence spectra in the Voigt geometry. (a) GaAs-2. The oscillations in photoluminescence intensity with field are 
    attributed to oscillations in magneto-absorption due to the diamagnetic exciton effect~\cite{ref:seisyan2012dee}, ${T=2~\text{K}}$, excitation and collection are performed in linear polarizations oriented at $\pm 45^\circ$ with respect to the magnetic field direction, 1 mW excitation power at 810~nm. 
    (b)~InP-2, ${T=2.3~\text{K}}$, $\sigma$-polarization excitation, all polarizations collected, 40~$\mu$W above band-gap excitation power. 
    (c) CdTe-2. ${T=1.6~\text{K}}$. $\pi$-polarization excitation,  $\sigma$-polarization collection, 20~$\mu$W above band-gap excitation power.
    }
    \label{fig:magnetoPL}
\end{figure}

\section{GaAs and CdTe \Tone measurements}
\label{appendix:GaAsCdTe} 

\begin{figure*}[ht]
\includegraphics[width=6.5in]{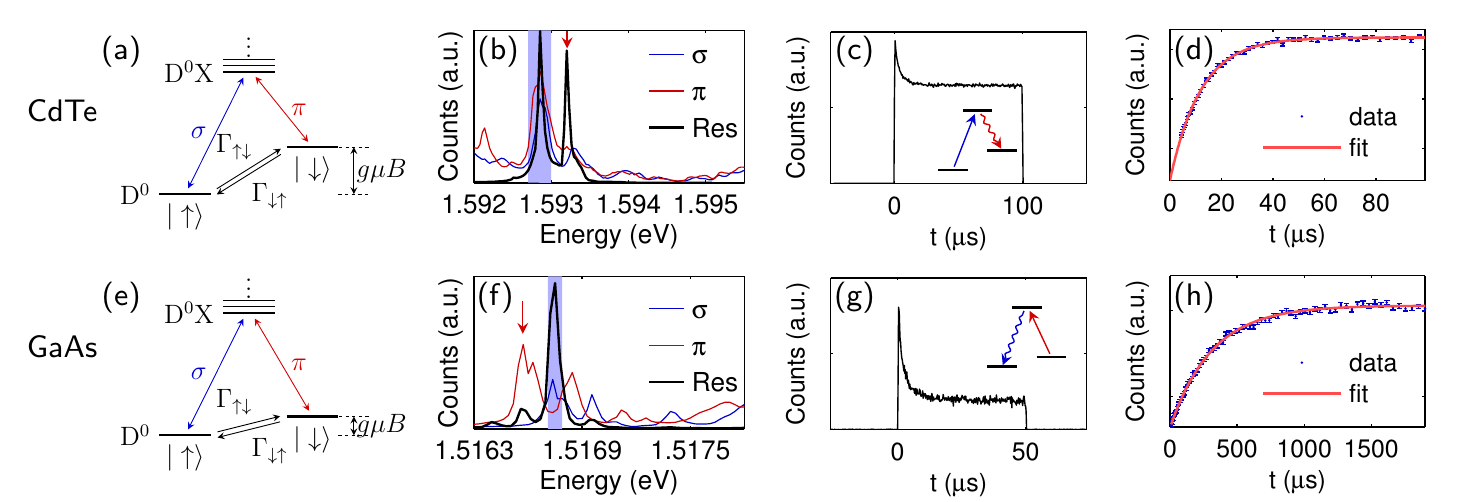}
\caption{(a) Energy level diagram for donor system in CdTe. (b) Photoluminescence spectrum of CdTe at ${B=3.5~\text{T}}$, ${T=1.5~\text{K}}$.  Excitation at 1.653~eV with 50~$\mu$W for the two above band spectra (red and blue). Excitation at 1.593 with 50~$\mu$W for resonant spectrum (black), as shown by the red arrow. (c) Optical pumping trace for CdTe at 3.5~T, 1.5~K. Power 50~$\mu$W. Laser pulse lasts 100~$\mu$s. (d) \Tone measurement for CdTe at 3.5~T, 1.5~K. Power 50~$\mu$W. ${T_1 = (12.0\pm0.2)~\mu\text{s}}$.  (e) Energy level diagram for donor system in GaAs. (f) Photoluminescence spectrum of GaAs at 7~T, 1.5 K. Excitation at 1.530~eV with 18~$\mu$W for the two above band spectra (red and blue). Excitation at 1.517 with 10~$\mu$W for resonant spectrum (black), as shown by the red arrow. (g) Optical pumping trace for GaAs at 7~T, 1.5~K. Power 10~$\mu$W.  Laser pulse lasts 50~$\mu$s. (d) \Tone measurement for GaAs at 7~T, 1.5~K. Power 10~$\mu$W.  ${T_1 = (313\pm5)~\mu\text{s}}$. All data are taken with an excitation spot size~${\sim 30~\mu\mathrm{m}}$.}
\label{fig:exp_CdTe_GaAs}
\end{figure*}

Representative energy diagrams, spectra, optical pumping traces, and \Tone recovery traces for CdTe and GaAs are shown in Fig.~\ref{fig:exp_CdTe_GaAs}.  For GaAs and InP, the lower energy Zeeman pair transition was used for optical pumping. Although this results in a weaker signal due to the lower thermal population in the higher electron spin level, the lower energy transition is clearly resolved from all other D$^0$X transitions enabling efficient optical pumping. For CdTe, there is significant inhomogeneous optical broadening of the D$^0$X lines. This can be observed by comparing the non-resonant and resonant excitation spectral linewidths in Fig.~\ref{fig:exp_CdTe_GaAs}(b). Optical pumping visibility is thus significantly smaller in this sample relative to GaAs and InP. Empirically we find the best signal-to-noise is obtained by pumping the high-energy Zeeman pair transition due to the significantly larger thermal population in the lower energy spin state. Due to the large $g$-factor in CdTe, the thermal population in the high energy state at 7~T and 1.5~K is only 0.6\%.

\section{Surface depletion effects}
\label{appendix:depletion}

In the GaAs and InP samples, a \(\mu\)s-scale time-dependent increase in luminescence was observed in all band edge PL after the start of an optical excitation pulse. The magnitude of this effect varied significantly between samples and depended on both the wavelength and intensity of the optical excitation. The effect was  greater in InP than in GaAs and was greater in lower doped samples. It did not significantly depend on emission wavelength. Free exciton, D$^+$X, D$^0$X, and A$^0$X transitions all behaved similarly. 

Figure~\ref{fig:brightening}a depicts a representative example of this effect in sample InP-2. In two experiments at 0~T and 1.6~K, the D$^0$X emission is detected during an excitation pulse. In the first experiment, the sample is excited with a 5~$\mu$W excitation pulse resonant with the D$^0$X transition. During the application of this pulse a small increase in optical emission at the beginning of the pulse can be observed on the microsecond time-scale (blue trace). This effect decreases with increasing field and increases with excitation power intensity. In the second experiment, we use a 5~$\mu$W excitation pulse with energy greater than the bandgap. A significant emission increase is observed.  Using a pulse sequence similar to the \Tone sequence [Fig.~\ref{fig:setup}(e)], we find the sample relaxes to its initial state on the timescale of 50 microseconds. For all \Tone measurements reported in the main manuscript, the resonant excitation power is always kept low enough so that this emission enhancement effect is negligible. 

Due to this effect, we are unable to obtain a \Tone measurement for donors in InP-1, the InP sample with the lowest donor concentration. Example optical pumping curves for this sample are shown in Fig.~\ref{fig:brightening}(b) at 4~T and 1.6~K. The blue trace shows data corresponding to the standard optical pumping pulse sequence depicted in Fig.~\ref{fig:setup}(c). The visibility is poor and in addition to the small optical pumping feature, we see an increase in the PL intensity after the initial optical pumping phase. For InP-1, this ``brightening'' effect is observed at all reasonable powers (i.e. powers for which we can obtain enough signal to reliably obtain a \Tone measurement) and the decay of this signal is the dominant contribution in \Tone pulse sequence measurements.

Additionally, we performed a two-pulse experiment where a 50~$\mu$s pulse with energy above the bandgap is applied to the sample until 5~$\mu$s before the the optical pumping pulse begins. The effect of the pre-pulse is dramatic (Fig.~\ref{fig:brightening}): the larger visibility can be attributed to the pre-pulse depolarizing bound electron spins. However we also note that the intensity in the optically pumped steady-state, near the end of the pulse, is flat and significantly larger in the pre-pulse case. This indicates that in terms of emission intensity, the sample has reached steady-state during the application of the pre-pulse. We attribute the brightening effect to the elimination of near-surface fields in the GaAs and InP samples under optical illumination~\cite{ref:yablonovitch1989bbf}. There is evidence that very small fields, on the order of V/cm, can substantially quench fluorescence~\cite{ref:bludau1976iie}. We do not observe this effect in CdTe, which is a true bulk sample rather than a few-micron-thick film.

\begin{figure}
    \includegraphics{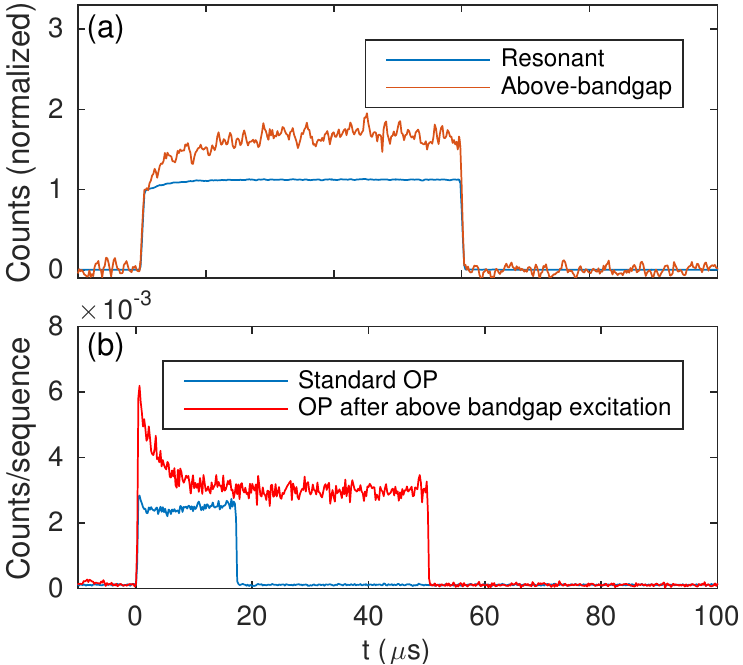}
    \caption{Time-resolved photoluminescence during optical pulse. (a) Collection of D$^0$X emission during resonant D$^0$X excitation (blue curve) and above-bandgap excitation (red curve). A significant increase in emission intensity is observed for above-bandgap excitation. (b) Optical pumping traces for InP-1 at 4~T. Blue curve: Standard optical pumping experiment. Red curve: Prior to the optical pumping pulse, a 50~$\mu$s long above-bandgap pre-pulse is applied. The end of the pre-pulse is 5~$\mu$s before the start of the optical pumping pulse.} 
    \label{fig:brightening}
\end{figure}


\section{Numerical solution of donor-bound electron in magnetic field}
\label{appendix:h2db}

\begin{figure}[htb]
\includegraphics{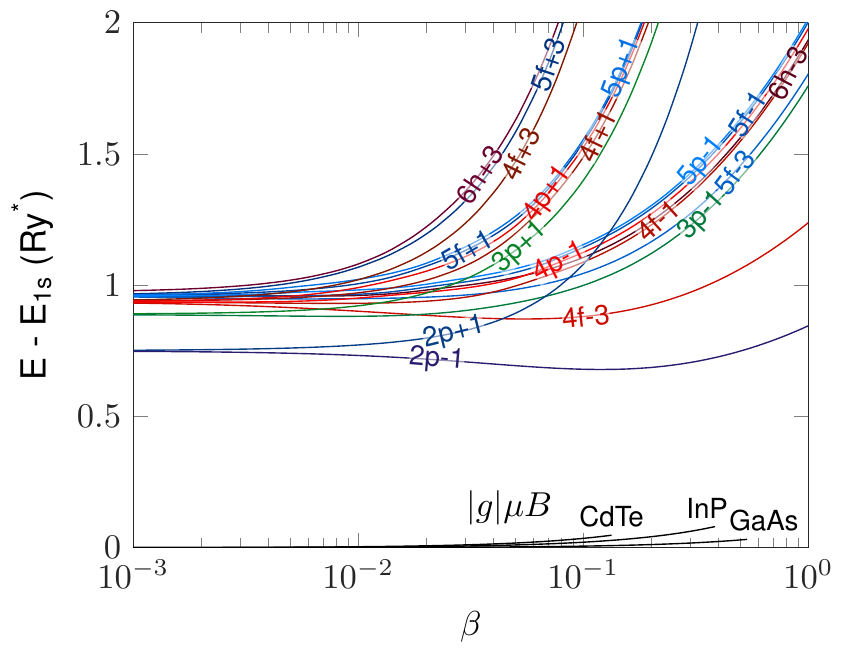}
\caption{Energies of excited state orbitals vs. dimensionless magnetic field  \(\beta\) from numerical simulation of hydrogen atom in magnetic field. The plot shows the energy difference of 18 excited states from the ground state with the same spin projection. States are labeled by their zero-field quantum numbers. Also plotted is the Zeeman splitting energy $|g|\mu B$ in units of effective Rydberg, the maximum $\beta$ of the plot being the maximum experimental $\beta$ obtained for each material. The Zeeman energy can be ignored compared to the orbital energy.}
\label{fig:numericenergies}
\end{figure}

The numerical solution of the hydrogen atom in a magnetic field is a nontrivial problem~\cite{toddkarin:Rosner1984}. Of particular difficulty is the transition from the low-field to high-field regime, where the solutions cannot be conveniently expanded in hydrogen or Landau orbitals~\cite{toddkarin:Li-Bo2009}.
We have used a readily available finite element solver to find the energies and wave functions of hydrogen in a magnetic field of arbitrary strength~\cite{toddkarin:Schimeczek2014}. These solutions can be mapped onto the donor-bound electron problem by replacing the electron mass, Bohr radius, \(g\)-factor and other parameters by their effective values for the donor-bound electron.
The magnetic field is measured by a dimensionless quantity \(\beta=B/B_0\), where the reference magnetic field \({B_0=2\hbar/[|e|(a_B^*)^2]}\) is found by considering when the Larmor radius \(\sqrt{2\hbar/|eB|}\) is equal to the donor Bohr radius.
For GaAs, InP, and CdTe, \(B_0\) is, 13.4~T, 19.3~T and 49.8~T respectively. In our experiment, the maximum applied field is 7~T, implying that the 1s wave function is a good approximation for the ground state. However, we note that for higher energy orbitals \(n\), the magnetic field at which magnetic effects begin to dominate Coulomb ones occurs at \(B_0/n^3\)~\cite{toddkarin:Schimeczek2014}. Thus, higher energy orbitals are significantly perturbed even at small \(\beta\). The energy difference between the excited states and the ground state are shown in Fig.~\ref{fig:numericenergies}. The energy is scaled by the effective binding energy \(\mathcal E_{Ry}^*\), which is 5.8, 7.0 and 13.6 meV for GaAs, InP and CdTe respectively. 

\section{Theory of spin-relaxation via the admixture mechanism}
\label{appendix:admixtureTheory}

In this appendix, we calculate the spin-relaxation rate due to the admixture mechanism in several different ways. First we provide general simplifications that are common to all calculations. We then evaluate the expression for \Tone numerically at all fields and analytically at low and moderate fields.

\subsection{General expression for the admixture spin-relaxation rate} 

It is convenient to represent the spin-relaxation rate, Eq.~\eqref{eq:T1start}, in a simplified form. First, we investigate which excited states may contribute to spin relaxation by symmetry.

The Dresselhaus spin-orbit interaction Hamiltonian~\eqref{eq:Hso} is cubic in the electron wavevector. It can be conveniently decomposed in the spherical angular harmonics, $Y_{l}^{m}({\theta}_{ k}, {\phi}_{k})$, where the subscripts $l=0,1,2,\ldots$, $m=-l,-l+1, \ldots,l-1,l$, and ${\theta}_{ k}$, ${\phi}_{ k}$ are the polar and azimuthal angles of the wavevector $\mathbf k$ in the spherical coordinate system with $z$ being the polar axis. Corresponding summands in Eq.~\eqref{eq:Hso} take the form:
\begin{subequations}
\label{third}
\begin{align}
&\frac{k_x(k_y^2 - k_z^2)}{k^3} = \sqrt{\pi}\left(\frac{Y_3^{3} - Y_3^{-3} }{\sqrt{35}}  + \frac{Y_3^1 - Y_3^{-1}}{\sqrt{21}} \right), \\
&\frac{k_y(k_z^2 - k_x^2)}{k^3} = \sqrt{\pi} \left(\frac{Y_3^{3} + Y_3^{-3} }{i\sqrt{35}}  - \frac{Y_3^1 + Y_3^{-1}}{i\sqrt{21}} \right),\\
&\frac{k_z(k_x^2-k_y^2)}{k^3}= 2\sqrt{\frac{2\pi}{105}}\left(Y_3^2+Y_3^{-2}\right).
\end{align}
\end{subequations}
Here the arguments of the spherical harmonics are omitted for brevity. In our frame of axes where ${\mathbf B \parallel z}$, the eigenstates of the donor-bound electron in the magnetic field are characterized by the angular momentum component $m$ onto the $z$ axis. Note that the term $ \sigma_zk_z(k_x^2-k_y^2)$ in Eq.~\eqref{eq:Hso} does not play a role in the spin flip process. Hence, the intermediate states for the admixture mechanism, in agreement with the first two lines of Eq.~\eqref{third}, are those with ${m=\pm 1}$, ${m=\pm 3}$. In relatively weak fields where the magnetic field does not perturb the ground and excited stated wave functions, the donor-bound electron has spherical symmetry and Eq.~\eqref{third} imposes a strict selection rule for the excited states: only ${l=3}$ (and ${m=\pm1}$, ${m=\pm 3}$) can cause spin-relaxation.

As such, in the sum over excited states in Eq.~\eqref{eq:M}, we only need to include \({m=\pm1}\) and \({m=\pm3}\) states. We also note that due to the azimuthal symmetry,
\[\langle \nu,\pi_z,m| k_x (k_y^2-k_z^2) | {\text{1s}} \rangle = e^{-i m \frac{\pi}{2}} \langle \nu,\pi_z,m| k_y (k_x^2-k_z^2) | {\text{1s}} \rangle,\]
it is sufficient to calculate the contribution due to the $\sigma_x k_x(k_y^2-k_z^2)$ term in the Dresselhaus Hamiltonian. By combining positive and negative \(m\) terms and simplifying, we find
\begin{align}
\label{eq:M:app}
& |M_{\uparrow\downarrow}|^2 = \frac{2\gamma^2 \hbar}{\rho \omega_{\mathbf q,\alpha}} |e A_{\mathbf q,\alpha} |^2 \cdot  \\
& \nonumber  \Biggl| \hspace{0.3cm}
 \sum_{\mathclap{\substack{m=1,-3\\ \nu=1,2,\dots\\ \pi_z=1}}} 
 \langle {\text{1s}} |e^{i {\mathbf q} \mathbf r} |\nu,\pi_z,m \rangle 
 \langle \nu,\pi_z,m| k_x (k_y^2-k_z^2) | {\text{1s}} \rangle G_{\nu;m} 
 \Biggr|^2
\end{align}
where \[G_{\nu;m} = (\Delta E_{\nu,\pi_z,m} - g \mu B)^{-1} - (\Delta E_{\nu,\pi_z,-m} + g \mu B)^{-1}, \] and \(\pi_z=1\) by symmetry.
By integrating over phonon modes, we find the general expression
\begin{widetext}
\begin{equation}\label{eq:T1generalAdm}
\frac{1}{T_1}= F_{ph} \frac{\gamma^2}{2 \pi^2 \hbar^2 \rho} \sum_\alpha \sum_{m=1,-3} \frac{|g \mu B|}{s_\alpha^3}  \int d \Omega_q |e A_{\mathbf q, \alpha}|^2  
  \Bigg| 
 \sum_{\nu} 
 \left[ \langle {\text{1s}} |e^{i \mathbf q \mathbf r} |\nu,\pi_z,m \rangle\right]_{q=q_\alpha}  \langle \nu,\pi_z,m| k_x (k_y^2-k_z^2) | {\text{1s}} \rangle G_{\nu;m} 
 \Bigg|^2\:,
\end{equation}
\end{widetext}
where the phonon matrix element is evaluated at a wavevector $q$ magnitude corresponding to the Zeeman energy ${q=q_\alpha \equiv |g\mu B|/\hbar s_\alpha}$. In the following sections, we will evaluate Eq.~\eqref{eq:T1generalAdm} using numerically calculated functions and an analytic approximation for the hydrogenic wavefunctions in a magnetic field.

\subsection{Numerical calculation of admixture spin-relaxation rate}
\label{appendix:admixtureNumerical}

For the two matrix elements in Eq.~\eqref{eq:T1generalAdm}, the integrals over the azimuthal angle of the position vector $\mathbf r$ can be performed analytically. This greatly speeds the evaluation time and improves the accuracy of the numerical calculation. The wave functions are written in cylindrical coordinates as \( \langle \mathbf r|\nu,\pi_z,m\rangle = \Phi_{\nu,\pi_z,m}(\rho,z) e^{i m \phi}\), where \(\rho\) is the radial coordinate, \(z\) the axial coordinate and \(\phi\) the azimuthal angle.  By transforming the differential operators into cylindrical coordinates and integrating over \(\phi\), we find 
\begin{align}
   \bra{\nu,\pi_z,\pm1}  k_x (k_y^2-k_z^2) & \ket{ \text{1s} }  = \nonumber \\
  -\frac{\pi}{4} \int \rho \, d\rho \, dz  \,  \Phi_{\nu,\pi_z,\pm1}&  \left[  \frac{1}{\rho^2}  \partial_\rho - \frac{1}{\rho} \partial_\rho^2 - \partial_\rho^3 + 4 \partial_\rho \partial_z^2 \right] \Phi_\text{1s}, \nonumber \\
   \bra{\nu,\pi_z,\pm3}  k_x (k_y^2-k_z^2) & \ket{ \text{1s} }  = \nonumber \\
  -\frac{\pi}{4} \int \rho \, d\rho \, dz  \,  \Phi_{\nu,\pi_z,\pm3}&  \left[  \frac{3}{\rho^2}  \partial_\rho - \frac{3}{\rho} \partial_\rho^2 + \partial_\rho^3 \right] \Phi_\text{1s}, \label{eq:dresselhausN}
  \end{align}
and that the matrix element is zero for any other excited state magnetic quantum number, as it must be by symmetry.

Similarly, for the matrix element \({\langle \nu,\pi_z,m | e^{i \mathbf q  \mathbf r} | \text{1s} \rangle }\), we note that aligning the \({\phi=0}\) plane along \(\mathbf q\) results in multiplying the integrand by a phase $\exp{(- i m \phi_{q})}$, where \( \phi_q\) is the azimuthal angle of the phonon wavevector $\mathbf q$.
The \(\phi\) integral can then be performed analytically with the help of a Bessel function identity. Using a few additional simplifications involving the \(z\)-parity of the wave functions, the matrix element becomes
 \begin{align}
 \label{eq:general:me}
 \langle  \nu, \pi_z, m | & e^{i \mathbf q  \mathbf r}  | \text{1s} \rangle  = \, 4\pi \,e^{-i m \phi_q}  \int_0^\infty \rho\, d\rho \int_{0}^\infty \, dz \,   \cdot \\
 \nonumber
& \Phi_{\nu,\pi_z,m}  \Phi_\text{1s} \cos(z\, q \cos\theta_q)  J_m(\rho \,q \sin \theta_q),
 \end{align}
where $\theta_q$ is the polar angle of the wavevector $\mathbf q$,
and \(J_m\) is the \(m^\text{th}\) Bessel function of the first kind.

Lastly, we evaluate the integral over the phonon azimuthal angle \(\phi_q\) in Eq.~\eqref{eq:T1generalAdm}. We additionally note that \({m=\pm1}\) states cannot interfere with \({m=\pm3}\) states due to the \(e^{-i m \phi_q}\) factor in Eq.~\eqref{eq:general:me}. We thus arrive at the expression for \Tone via the admixture mechanism,
\begin{widetext}
\begin{equation}\label{eq:generalAdmixture}
    \frac{1}{T_{1}} =   F_{ph} \frac{\gamma^2  |g \mu B|}{2 \pi \hbar^2 \rho}  \int \sin \theta_q d \theta_q \sum_\alpha \sum_{m=1,-3}   \left|\sum_{\nu=1}^\infty  \left[  \langle  \text{1s} | e^{i \mathbf q  \mathbf r}  | \nu, \pi_z, m \rangle \right]_{\phi_q=0}^{q=q_\alpha}     \bra{\nu,\pi_z,m}  k_x (k_y^2-k_z^2)  \ket{\text{1s} } \right|^2 P_\alpha(\theta_q)\:,
\end{equation}
\end{widetext}
where the phonon matrix element is evaluated at the wavevector corresponding to the Zeeman energy and at \({\phi_q=0}\)~\footnote{Since \(\phi_q\) just adds a phase, the expression is valid for any choice of \(\phi_q\).}. The functions $P_\alpha(\theta_q)$ describe the contributions of different phonon modes and different electron-phonon interaction mechanisms. We take into account both the piezoelectric interaction with longitudinal and transverse modes, Eq.~\eqref{U:piezo}, as well as the deformation potential interaction. The latter is described by the Hamiltonian~\cite{toddkarin:Gantmakher1987,PhysRevB.72.125340} 
\begin{equation}
\label{U:def}
U_{ph}^{(dp)}(\mathbf r) = \sqrt{\frac{\hbar}{2\rho\omega_{\mathbf q\alpha}}} e^{i \mathbf q \mathbf r} q (\xi_i \hat e_i) D b^\dag_{\mathbf q\alpha} + {\rm c.c.},
\end{equation}
where $D$ is the deformation potential constant and involves longitudinal phonons only. As a result, disregarding the interference of piezo and deformation potential interactions, we have
\begin{equation}
\begin{aligned}
P_1(\theta_q) & = \frac{9e^2 h_{14}^2}{s_l^3} \cos^2 \theta_q \sin^4 \theta_q\:, \\
P_2(\theta_q) & =\frac{e^2 h_{14}^2}{8s_t^3} (27 + 28 \cos 2 \theta_q + 9 \cos 4 \theta_q )\:, \\
P_3(\theta_q) & = \frac{ 2(g \mu B)^2 D^2}{\hbar^2 s_l^5} .
\end{aligned}
\end{equation}
The simplified matrix elements Eq.~\eqref{eq:dresselhausN}, \eqref{eq:general:me} are calculated numerically using standard procedures. The scripts have been made readily available~\cite{h2dblab}.

\subsection{Admixture mechanism in moderate magnetic fields}
\label{appendix:admixtureModerateField}

In the regime of moderate fields, when ${l_b \sim a_B^*}$, we take into account the modification of the excited state wavefunctions by the magnetic field. In this regime, admixture with the lowest energy excited states (p-shell states) is allowed and we will further take into account only these two states assuming a Gaussian form of the wavefunctions:
\begin{eqnarray}
\label{eq:app_wfs}
\psi_{\text{1s}} &=& \frac{1}{\pi^{3/4} l_{\rho} \sqrt{l_z}} \ \exp{\left(-\frac{x^2+y^2}{2 (l_{\rho,\text{1s}})^2} -\frac{z^2}{2 (l_{z,\text{1s}})^2}\right)},  \\
\psi_{\text{2p}_\pm} &=& \frac{ (x \pm iy)}{\pi^{3/4} \chi^{5/2} l_\rho^2 \sqrt{l_z}} \ \exp{\left(-\frac{x^2+y^2}{2 (l_{\rho,\text{2p}})^2}-\frac{z^2}{2 (l_{z,\text{2p}})^2}\right)}\:. \nonumber
\end{eqnarray}
Further, for simplicity, we will assume a proportionality ${l_{\rho,\text{2p}} = \chi l_{\rho,\text{1s}} }$ and ${l_{z,\text{2p}} = \chi l_{z,\text{1s}} }$ for the analytic wavefunctions. Here ${l_{\rho,\text{1s}}=[1/(a_B^*)^2+1/{(2l_b^2)}]^{-1/2}}$ and ${l_{z,\text{1s}} = a_B^*}$ are the wave function effective sizes in the $(xy)$-plane and the $z$-direction, respectively.
At zero field, $l_{\rho,\text{1s}}$ is just the Bohr radius and both lengths coincide. A non-zero magnetic field shrinks the wave function in the $(xy)$-plane leading to anisotropy of the $\text{1s}$-state with ${l_{\rho,\text{1s}} < l_{z,\text{1s}}}$. In the limit of very strong fields ${l_{\rho,\text{1s}}=\sqrt{2}l_b}$, in agreement with the free-electron wave function in a magnetic field in the symmetric gauge.
The exact numerical wavefunctions (Appendix~\ref{appendix:h2db}) are used to determine the value of \(\chi\) for the analytic wavefunctions, Eq.~\eqref{eq:app_wfs}. To determine the best value of \(\chi\), we numerically optimized the overlap integral between the analytic wavefunctions, Eq.~\eqref{eq:app_wfs}, and the numerical ones.
The ratio of the wavefunction size for the 2p$_{-1}$ and 1s states are shown in Fig.~\ref{fig:numericChi}. In the figure we have also shown the limiting values of \(\beta\) for which the experimental high-field dependence is observed. By averaging over the ratio of lengths for \(\rho\) and \(z\) directions, we obtain a best-choice \(\chi\) of 1.5, 1.7 and 2.2 for GaAs, InP, and CdTe respectively.

\begin{figure}[htb]
\includegraphics{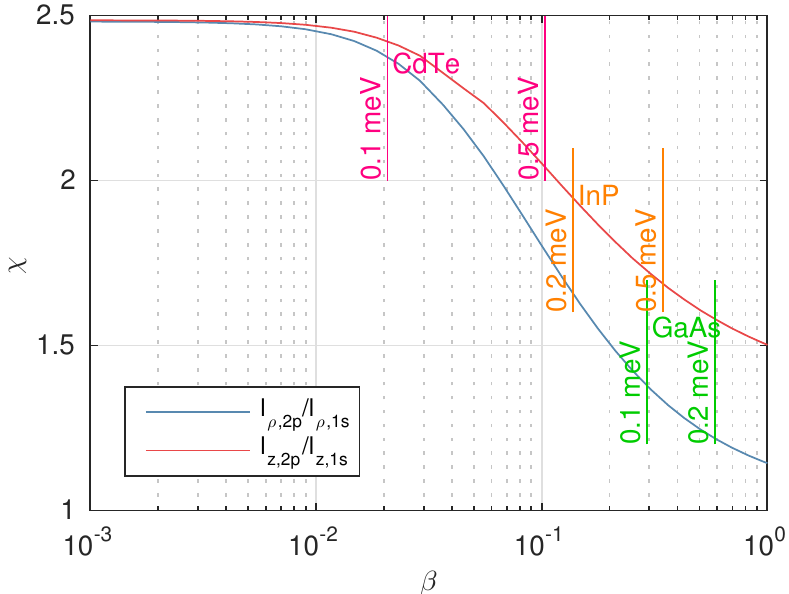}
\caption{
The overlap integral between the Gaussian approximation for the wavefunction, Eq.~\eqref{eq:app_wfs}, and the numerical solution was maximized as a function of \(\beta\). The ratio of the radial and axial Gaussian sizes is plotted as a function of \(B\). Also shown are the experimental limits of the ``high-field" regime where \(T_1\) goes as {$B^{-\nu}$} for the three different materials. Using these limits, reasonable choices of \(\chi\) are 1.5, 1.7 and 2.2 for GaAs, InP, and CdTe respectively.
}
\label{fig:numericChi}
\end{figure}

To obtain the matrix element $M_{\downarrow \uparrow}$ given by Eq.~\eqref{eq:M}, we need to calculate ${\langle 2 p_\pm | \partial_x(\partial_y^2-\partial_z^2)| \text{1s}\rangle}$ and ${\langle \text{1s} | e^{i \mathbf q \mathbf r}| \text{2p}_\pm \rangle}$.
Utilizing the Gaussian wave functions we assumed, Eq.~\eqref{eq:app_wfs}, the results for the integrals are
\begin{subequations}
\label{mat:dress}
\begin{eqnarray}
\langle \text{2p}_\pm | \partial_x(\partial_y^2-\partial_z^2)| \text{1s}\rangle &=& \frac{{\sqrt{2}\chi^{5/2}}}{{(1+\chi^2)^{7/2}}l_b^2 l_{\rho}}\:, \\
\langle \text{2p}_\pm | \partial_y(\partial_z^2-\partial_x^2)| \text{1s}\rangle &=& \pm \frac{i{\sqrt{2}\chi^{5/2}}}{{(1+\chi^2)^{7/2}} l_b^2 l_{\rho}}\:{.}
\end{eqnarray}
\end{subequations}
and
\begin{multline} \label{eq:eiqr_int1}
\langle \text{1s} | e^{i \mathbf q \mathbf r}| 2 p_\pm \rangle = { i 2\sqrt{2} \left( \frac{\chi}{1+\chi^2} \right)^{5/2}} (q_x \pm i q_y) l_\rho \times \\
\times {
\exp{\left(-\frac{\chi^2}{2(1 + \chi^2)} [(q_x^2+q_y^2) l_\rho^2+ q_z^2 l_{z}^2]\right)}} \:.
\end{multline}
Note that nonzero matrix elements in Eq.~\eqref{mat:dress} are proportional to $B$, so that they vanish in the limit of low fields (this regime is considered below in Appendix~\ref{appendix:admixtureLowField}).
As the magnetic fields in our experiments are not very strong, the change of the characteristic length is small and we can neglect the difference between $l_\rho$ and $l_z$ in the exponential part of Eq.~\eqref{eq:eiqr_int1} by setting ${l_{z,\text{1s}} = l_{\rho,\text{1s}} = l}$. In fact, the wavefunction size \(l\) is important only when the long-wavelength approximation for the electron-acoustic phonon interaction fails, in which case the result may be quite sensitive to the overall shape of the wave function, and, additionally, the deformation-potential interaction may be important.

Substituting these into Eqs.~\eqref{eq:T1start},\eqref{eq:M}, the relaxation rate is
\begin{multline}
\frac{1}{T_1} = {\frac{32}{\pi \hbar \rho} \frac{\chi^{10}}{(1 + \chi^2)^{12}}}  \frac{\gamma^2 (e h_{14})^2}{l_b^4} \left(\frac{1}{\Delta E_{R}}-\frac{1}{\Delta E_{R}+ \hbar \omega_c}\right)^2  \\ \times F_{ph} 
 \sum \limits_\alpha \frac{q_\alpha^3}{s_\alpha^2} {\exp{\left(-{\frac{\chi^2}{1 + \chi^2}} q_\alpha^2 l^2\right)}} \left \langle \sin^2 \theta_{ q} a_{\mathbf q, \alpha}^2 \right \rangle_{\Omega}\:{.}
\label{eq:T1_calc}
\end{multline}
The quantity $\Delta E = E_{\text{2p}_-} - E_{\text{1s}}$. Numerically we find that $\Delta E = 3/4 \mathcal E_{Ry}^*$
is a good approximation across the entire experimental range of fields (see Fig.~\ref{fig:numericenergies}).

For longitudinal phonons we have
\begin{equation}
\left \langle \sin^2 \theta_{ q} a_{ q, l}^2 \right \rangle_{\Omega} = \frac{8}{35}\:,
\end{equation}
while for transverse modes
\begin{equation}
\left \langle \sin^2 \theta_{ q} a_{ q, t}^2 \right \rangle_{\Omega} = \frac{32}{105}\:.
\end{equation}
Substituting these integrals into the Eq.~\eqref{eq:T1_calc}, the final result for the spin relaxation rate is
\begin{multline}
\label{Tone:adm:appendix}
\frac{1}{T_1} = {\frac{256 \chi^{10}}{35(1+\chi^2)^{12}}}\frac{\gamma^2 e^4 h_{14}^2 (g \mu)^3 B^5}{ \pi \rho \hbar^6} \times \\
\times
\left(\frac{1}{\Delta E_{R}}-\frac{1}{\Delta E_{R}+ \hbar \omega_c}\right)^2 F_{ph} \times \\
\times \left[ \frac{1}{s_l^5} {\exp{\left(-{\frac{\chi^2q_l^2 l^2}{1 + \chi^2}} \right)}} + \frac{4}{3s_t^5} {\exp{\left(-{\frac{\chi^2 q_t^2 l^2}{1 + \chi^2}} \right)}} \right] \:,
\end{multline}
in agreement with Eq.~\eqref{Tone:adm:analyt} of the main text.



We note that in the limit of very strong magnetic fields where ${l_b \ll a_B^*}$ and ${\hbar\omega_c \gg \Delta E_R}$ one has to take into account the modification of the separation between the ground and the excited states by the magnetic field. As a rough estimate one may replace $\Delta E_R$ in Eq.~\eqref{Tone:adm:appendix} by $\hbar\omega_c$, in which case, within the LWA, one has ${T_1 \propto B^3}$. Our estimates show that this limit is not fulfilled in any sample for the magnetic fields under study.

Finally, we briefly analyze the deformation potential interaction in which case instead of Eq.~\eqref{U:piezo} one has {Eq.~\eqref{U:def}.} It follows from Eq.~\eqref{U:def} that transition can be assisted by the logitudinal acoustic phonons only. Making use of the analytical form of the wavefunctions, Eq.~\eqref{eq:app_wfs}, and the matrix elements of the Dresselhaus spin-orbit interaction, Eq.~\eqref{mat:dress}, as well as Eq.~\eqref{eq:eiqr_int1} we obtain
\begin{multline}
\label{t1:def:inter}
\frac{1}{T_1} = \frac{{64} \chi^{10}}{3\pi {\hbar} \rho(1+\chi^2)^{12}} \left(\frac{g\mu B}{\hbar s_l}\right)^5  \frac{{\gamma^2}D^2}{{s_l^2}l_b^4} F_{ph}\times
\\
\left(\frac{1}{\Delta E_{R}}-\frac{1}{\Delta E_{R}+ \hbar \omega_c}\right)^2 \exp{\left(-\frac{\chi^2 q_l^2l^2}{1+\chi^2}\right)}.
\end{multline}
The angular integrations over the phonon wavevectors has been carried out, as before, neglecting the difference between $l_{\rho,\text{1s}}$ and $l_{z,\text{1s}}$ in the exponent and using the expression ${\langle (\xi_i \hat e_i)^2 (\xi_x^2+\xi_y^2) \rangle_\Omega=2/3}$. The analysis shows that the deformation potential contribution is much smaller than the piezo-interaction, Eq.~\eqref{Tone:adm:appendix} for the relevant magnetic fields. The contribution from the deformation potential interaction is more important than the piezo-interaction at high fields due to its stronger $B$-field dependence. The crossover field for GaAs is about 40~T, which is much larger than the magnetic fields in this study. The crossover fields for InP and CdTe are 9.1~T and 3.9~T. Although deformation potential interaction for these two materials is comparable to the piezo-interaction at the fields achievable in our experiment, it is still much weaker than the direct spin-phonon interaction.

\subsection{Admixture mechanism in low magnetic fields}
\label{appendix:admixtureLowField}

In the low-field limit we take the wave functions of the ground 1s and excited ($|e \rangle = |n\text{f},m \rangle$, where f denotes f-orbitals) states in a hydrogen-like form:
\begin{eqnarray}
\psi_{\text{1s}} &=& \frac{1}{\sqrt{\pi a^3}} e^{-r/a} \nonumber \:, \\
\psi_{n\text{f},m} &=& R_{n3}(r) Y_3^m ({\theta}, {\phi})\:.
\end{eqnarray}
{Here 
and $R_{n3}$ are the radial functions of the f-orbitals ($l = 3$). }

To calculate the spin-flip rate we use the matrix element $M_{\downarrow \uparrow}$ given by Eq.~\eqref{eq:M}. We note that in the absence of a magnetic field the following relation for the matrix elements of $H_{so}$ holds
\begin{equation}
\left \langle \text{1s} \uparrow \left| H_{so} \right| n\text{f}, m \downarrow \right \rangle = - \left \langle n\text{f}, -m \uparrow \left| H_{so} \right| \text{1s} \downarrow \right \rangle \:.
\end{equation}
Using this relation and keeping in mind that the energies of $m$ and $-m$ states are the same at zero magnetic field the second order matrix element, Eq.~\eqref{eq:M}, vanishes as ${B \to 0}$~\cite{toddkarin:Khaetskii2001}. At nonzero magnetic field $M_{\downarrow \uparrow}$ becomes nonzero due to (i) Zeeman splitting of spin sublevels and (ii) orbital splitting of ${m = \pm 3}$ and ${m = \pm 1}$ states. Taking into account only the Zeeman splitting we obtain the following expression for the matrix element $M_{\downarrow \uparrow}$ in the low-field limit (the effect of the field-induced orbital splitting is briefly addressed in the end of this subsection)
\begin{equation}
\label{eq:app_M_lf}
M_{\downarrow\uparrow} = 2g\mu B \sum_{n,m} \frac{\langle \text{1s}, \downarrow |U_{ph}|n\text{f},m, \downarrow \rangle \langle n\text{f},m,\downarrow |H_{so}|\text{1s}, \uparrow \rangle}{(E_{n\text{f},m} - E_{\text{1s}})^2}\:.
\end{equation}
Here the wavefunctions and energies are taken at $B = 0$.
The sum over $m$ in Eq.~\eqref{eq:app_M_lf} can be evaluated using the fact that at ${B = 0}$ the energy spectrum is degenerate with respect to $m$ and the following formula:
\begin{multline}
\label{eq:app_sum_m}
\sum \limits_{m} \langle \text{1s}, \downarrow |e^{i \mathbf q \mathbf r}|n\text{f},m, \downarrow \rangle \langle n\text{f},m,\downarrow |H_{so}|\text{1s}, \uparrow \rangle = \\
= -4 \pi \frac{\gamma}{a^3} \langle R_{n3}(r) \Phi(r) \rangle_r \langle \psi_{\text{1s}}(r) R_{n3}(r) j_3(qr) \rangle_r F(\bm \xi)\:.
\end{multline}
Here ${\Phi = (1 + 3a/r + 3a^2/r^2) \psi_{\text{1s}}}$, $j_3$ is the spherical Bessel function of the third order, $F(\bm \xi) = \xi_x(\xi_y^2 - \xi_z^2) + i \xi_y(\xi_z^2 - \xi_x^2)$, and ${\bm \xi = \mathbf q /q}$. Angular brackets with the subscript $r$ denote integration over $r$, i.e. $\langle f_1(r) f_2(r) \rangle_r = \int r^2 f_1 f_2 dr$. Equation~\eqref{eq:app_sum_m} is obtained using the decomposition of $e^{i \mathbf q \mathbf r}$ over the spherical harmonics, Eqs.~\eqref{third} and orthogonality of $Y_l^m$ with respect to $m$.

The spin-flip rate calculated according to Eqs.~\eqref{eq:T1start} and \eqref{eq:app_M_lf} has a form:
\begin{equation}
\label{eq:app_T1_lf}
\frac{1}{T_1} = \frac{32 \pi}{105^2} \frac{(\gamma e h_{14})^2}{\rho \hbar^8 (\mathcal E_{Ry}^*)^4} S^2 (g \mu B)^9 \sum \limits_\alpha \frac{1}{s_\alpha^9} \left \langle |F(\bm \xi)|^2 a_{\mathbf q, \alpha}^2 \right \rangle_{\Omega}\:,
\end{equation}
where $a_{\mathbf q, \alpha} = A_{\mathbf q, \alpha}/h_{14}$, $\langle \dots \rangle_\Omega$ denotes an average over the angles ${\theta}_{ q}$ and ${\phi}_{ q}$ of $\mathbf q$, and
\begin{equation}
S =  \sum \limits_{n \geq 4} \frac{1}{\left( 1 - E_{n\text{f}}/\mathcal E_{Ry}^* \right)^2} \left \langle R_{n3} \Phi \right \rangle_r \left \langle \psi_{\text{1s}} \bar{r}^3 R_{n3} \right \rangle_r\:
\end{equation}
is the dimensionless sum over all excited states with different $n$ (including the states of the continuous spectrum) and ${\bar{r} = r/a_B^*}$.
In the evaluation of Eq.~\eqref{eq:app_T1_lf} we used the long wavelength approximation for phonons by using the asymptotic $j_3(qr) \approx (qr)^3/105$ at $qr \ll 1$.

For longitudinal acoustic phonons, ${\hat{\mathbf e}^{(\mathbf{q},l)} = \bm \xi}$, and hence
\begin{equation}
\left \langle |F(\bm \xi)|^2 a_{\mathbf q, \alpha}^2 \right \rangle_{\Omega} = \frac{96}{5005}\:.
\end{equation}
For transverse acoustic modes, the polarization vectors satisfy ${\langle \hat{e}_i^{(\mathbf{q},t)} \hat{e}_j^{(\mathbf{q},t)} \rangle = (\delta_{ij}-\xi_i \xi_j)/2}$. Considering that there are two transverse modes, we obtain
 \begin{equation*}
\left(\sum_{i, j, k} \beta_{i j k} \xi_i \xi_j \hat{e}_k^{(\mathbf{q},t)}\right)^2 = 4(\xi_x^2\xi_y^2+\xi_x^2\xi_z^2+\xi_y^2\xi_z^2-9\xi_x^2\xi_y^2\xi_z^2)\:,
\end{equation*}
and
\begin{equation}
\left \langle |F(\bm \xi)|^2 a_{\mathbf q, \alpha}^2 \right \rangle_{\Omega} = \frac{2048}{45045}\:.
\end{equation}
Using these averages one finally obtains for the spin relaxation rate
\begin{equation}
\label{eq:app_T1_final}
\frac{1}{T_1} = \zeta \frac{(\gamma e h_{14})^2}{\rho \hbar^8 (\mathcal E_{Ry}^*)^4} S^2 (g \mu B)^9 \left[ \frac{1}{s_l^9} + \left( \frac43 \right)^3 \frac{1}{s_t^9} \right]\:,
\end{equation}
with $\zeta = 1024 \pi /(1287 \times 35^3) \approx 0.000175$.

Let us now turn to the evaluation of the $S$ parameter. It comprises the sum over the discrete spectrum (index $n$) and the integral over the continuum spectrum (index $\eta$):
\begin{multline}
\label{eq:app_S}
S = \sum \limits_{n = 4}^{+\infty} \frac{1}{(1 - 1/n^2)^2} \left \langle R_{n3} \Phi \right \rangle_r \left \langle \psi_{\text{1s}} \bar{r}^3 R_{n3} \right \rangle_r + \\
+ \int \limits_0^{+\infty} \frac{\left \langle R_{\eta3} \Phi \right \rangle_r \left \langle \psi_{\text{1s}} \bar{r}^3 R_{\eta3} \right \rangle_r}{(1 + \eta^2)^2} d\eta\:.
\end{multline}
Matrix elements entering Eq.~\eqref{eq:app_S} are calculated analytically using formulas (f,1) and (f,2) of Ref.~\cite{toddkarin:LandauQM}:
\begin{align}
\label{eq:app_I1}
\left \langle \psi_{\text{1s}} \bar{r}^3 R_{n3} \right \rangle_r = \frac{96}{\sqrt{\pi} n^5} \sqrt{\frac{(n+3)!}{(n-4)!}} \frac{(1-1/n)^{n-5}}{(1+1/n)^{n+5}}\:, \nonumber \\
\left \langle \psi_{\text{1s}} \bar{r}^3 R_{\eta3} \right \rangle_r = \frac{96}{\sqrt{\pi}} \left( \frac{\eta}{1 - e^{-2\pi/\eta}} \right)^{1/2} \times \nonumber \\
\sqrt{\prod \limits_{s=1}^3 (s^2 \eta^2 +1)} \frac{(1- i \eta)^{-i/\eta-5}}{(1+ i \eta)^{-i/\eta+5}}\:,
\end{align}
and
\begin{align}
\label{eq:app_I2}
 \left \langle R_{n3} \Phi \right \rangle_r =  \frac{16}{7! \sqrt{\pi} n^5} \sqrt{\frac{(n+3)!}{(n-4)!}} \times \nonumber \\
 \sum \limits_{\nu = 4,5,6} c_\nu \frac{(\nu - 1)!}{(1 + 1/n)^\nu} {}_2{F}_1 \left(4-n, \nu, 8, \frac{2}{n+1}\right)\:, \nonumber \\
  \left \langle R_{\eta3} \Phi \right \rangle_r =  \frac{16}{7! \sqrt{\pi}} \left( \frac{\eta}{1 - e^{-2\pi/\eta}} \right)^{1/2} \sqrt{\prod \limits_{s=1}^3 (s^2 \eta^2 +1)} \times \nonumber \\
  \sum \limits_{\nu = 4,5,6} c_\nu \frac{(\nu - 1)!}{(1 + i \eta)^\nu} {}_2{F}_1\left(\frac{i}{\eta} + 4, \nu, 8, \frac{2 i \eta}{1+ i \eta}\right)\:.
\end{align}
Here ${c_4 = 3}$, ${c_5 = 3}$, ${c_6 = 1}$, and ${}_2{F}_1$ is the ordinary hypergeometric function. Using these matrix elements, a numerical summation in Eq.~\eqref{eq:app_S} is performed yielding ${S \approx 0.487}$. To analyze the contribution of the excited states we provide two estimates for $S$: ${S_{\mathrm{low}}< S < S_{\mathrm{up}}}$. Here the lower limit ($S_{\mathrm{low}}$) is the first term in the sum with $n = 4$, and the upper limit ($S_{\mathrm{up}}$) is the sum over a complete set of functions $\{ R_{n3}, R_{\eta 3} \}$ with a fixed denominator equal to an energy distance between the 4f and 1s state:
\begin{equation}
S_{\mathrm{low}} = \frac{256}{225} \left \langle \mathrm R_{43} \Phi \right \rangle_r \left \langle \psi_{\text{1s}} \bar{r}^3 \mathrm R_{43} \right \rangle_r \approx 0.006 \:,
\end{equation}
\begin{equation}
S_{\mathrm{up}} = \frac{256\left \langle \Phi \bar{r}^3 \psi_{\text{1s}} \right \rangle_r }{225} \approx 1.9\:.
\end{equation}
Noteworthy, $S$ exceeds $S_{\mathrm{low}}$ by more than two orders of magnitude, demonstrating the importance of accounting for all excited states of the spectrum. However, for highly excited states the LWA breaks down which somehow reduces the estimate of $S$. Moreover, for the experimental donor densities the overlap of states with large ${n \gtrsim 5}$ belonging to different donors is not negligible. The account for such an overalap is beyond the scope of the present paper.

Equation~\eqref{eq:app_T1_final} was derived assuming that the Zeeman splitting dominates the orbital $B$-linear splitting of the excited states with opposite values of $m$. Such an approximation works well in quantum dot systems where the dot anisotropy lifts the degeneracy in $m$. For the donor-bound electron this is not the case, since the problem (at ${B=0}$) has a spherical symmetry. In this situation, the splitting of the $n\text{f}$ states with $m$ and $-m$ is $|m|\hbar\omega_c \gg g\mu B$. To estimate \Tone in this case one should replace $(g\mu B)^9$ in Eq.~\eqref{eq:app_T1_final} by ${(g\mu B)^7 (\hbar\omega_c)^2}$ in agreement with Eq.~\eqref{lowlowB9} in the main text.

\section{Spin relaxation via the direct spin-phonon mechanism}\label{appendix:direct}

\subsection{General expression for the direct spin-phonon spin-relaxation rate}

The direct spin-phonon interaction Hamiltonian ~\cite{toddkarin:Dyakonov1986,toddkarin:Pikus1988,toddkarin:Frenkel1991} is 
\begin{equation}
    U_{dir} = \frac{\hbar v_0}{2}  [\sigma_x (u_{xy}k_y - u_{xz}k_z) + \sigma_y (u_{yz}k_z - u_{xy}k_x)],
    \label{eq:Uph}
\end{equation}
where we ignore the $\sigma_z$ term because it does not contribute to spin relaxation, ${\mathbf k = -i \bm \nabla - (e/\hbar) \mathbf A}$, and we use the symmetric gauge ${\mathbf A =  (-By/2, Bx/2, 0)}$. 
The deformation tensor $u_{ij}$ due to phonon \(\mathbf q, \alpha\) is
\begin{equation}
 u_{ij}^{\mathbf q, \alpha} = \sqrt{\frac{\hbar}{2\rho \omega_{\mathbf q,\alpha}}} e^{i (\mathbf q \mathbf r - \omega_{\mathbf q, \alpha} t)} \frac{i (\hat{e}_i^{(\mathbf q, \alpha)} q_j +  \hat{e}_j^{(\mathbf q, \alpha)} q_i)}{2} b^\dag_{\mathbf q, \alpha} + {\rm c.c.}
\end{equation}
where $\bm e_{\alpha}$ is the polarization of phonon mode $\alpha$:
\begin{equation}
\label{e:vec}
\begin{aligned}
\mathbf e_{l} &= q^{-1} [q_x,q_y,q_z],\\
\mathbf e_{t_1} &= (q_x^2+q_y^2)^{-1/2} [q_y,-q_x,0], \\
\mathbf e_{t_2} &= q^{-1}(q_x^2+q_y^2)^{-1/2} [q_xq_z,q_yq_z,-(q_x^2+q_y^2)]{.}
\end{aligned}
\end{equation}
Subscripts $t_1$ and $t_2$ denote two degenerate transverse modes.
The relaxation rate $\Gamma_{\downarrow \uparrow}$ is found from Eq.~\eqref{eq:M} using
\begin{equation}
\label{eq:app_M_dir}
M_{\downarrow \uparrow} = \left \langle \text{1s}, \downarrow \left| U_{dir} \right| \text{1s}, \uparrow \right \rangle\:.
\end{equation}
According to the general principles of quantum mechanics, the momentum operator and deformation tensor in Eq.~\eqref{eq:Uph} must be symmetrized, i.e. ${u_{ij} k_l \rightarrow \{u_{ij},k_l\}}$, where ${\{a,b\} = (ab+ba)/2}$~\cite{toddkarin:Khaetskii2001,toddkarin:Winkler2003,toddkarin:Ivchenko1997}. Due to this symmetrization and the fact that the ground state is a localized state, all terms with $k_z$ integrate to zero. Further simplifications yield 
\begin{align}
\nonumber \langle \text{1s},\downarrow | U_{dir} | \text{1s},\uparrow\rangle = i \frac{\hbar v_0}{4} \sqrt{\frac{\hbar}{2 \rho \omega_{\mathbf q, \alpha}}} \left( \hat{e}_{x}^{(\mathbf q, \alpha)} q_y +  \hat{e}_{y}^{(\mathbf q, \alpha)} q_x \right) \times  \\ 
\nonumber \left[ \langle \text{1s} | \{ \exp(i \mathbf{q r}), k_y \} | \text{1s} \rangle -i \langle \text{1s} | \{ \exp(i \mathbf{q r}), k_x \} | \text{1s} \rangle \right]
\label{eq:Uph_int},
\end{align}
\begin{eqnarray}
\langle \text{1s} | \{\exp(i \mathbf{q r}), k_x\} | \text{1s} \rangle &=& \frac{e B}{2 \hbar} \langle \text{1s} | \exp(i \mathbf{q r}) y | \text{1s} \rangle\:, \label{me:direct:gauge} \\
\langle \text{1s} | \{\exp(i \mathbf{q r}), k_y\} | \text{1s} \rangle &=& -\frac{e B}{2 \hbar} \langle \text{1s} | \exp(i \mathbf{q r}) x | \text{1s} \rangle\: \nonumber.
\end{eqnarray}

Substituting Eq.~\eqref{eq:app_M_dir} into Eq.~\eqref{eq:T1start} and taking the phonon factor $F_{ph}$ into consideration, we obtain the general expression for the spin-relaxation rate
\begin{multline}
\frac{1}{T_1} = F_{ph} \frac{v_0^2 e^2 B^2}{2^{9}  \pi^2 \hbar  \rho} \sum_{\alpha} \frac{q_{\alpha}^3}{s_{\alpha}^2} \int\, d\Omega_{q}\, \big|\left( \hat{e}_{x}^{\mathbf q, \alpha} \xi_y +  \hat{e}_{y}^{\mathbf q, \alpha} \xi_x \right)\cdot \\
 \langle \text{1s} | \exp(i \mathbf{q}_{\alpha} \mathbf{r}) (x-iy) | \text{1s} \rangle \big|^2\:,
\end{multline}
which can be evaluated either numerically or using an analytic approximation.

\subsection{Numerical calculation of direct spin-phonon spin-relaxation rate} 

Similar to Appendix~\ref{appendix:admixtureNumerical}, the azimuthal part of the integral \(\langle \text{1s} | \exp(i \mathbf{q}_{\alpha} \mathbf{r}) (x-iy) | \text{1s} \rangle\) can be calculated analytically to simplify the numerical calculation. We introduce the notation for this matrix element: 
\[
{\kappa}_\alpha(\theta_q) =  e^{i \phi_q}\langle  \text{1s}| e^{i \mathbf q_{\alpha} \mathbf r} (x - iy)  | \text{1s} \rangle,
\]
 and obtain 
\begin{multline}
\kappa_\alpha (\theta_{ q})  = \, 4\pi  \int_0^\infty \rho^2\, d\rho \int_{0}^\infty \, dz \, \Phi_{\text{1s}}^2(\rho,z)  \cdot \\
  \cos(z\, q_{\alpha} \cos\theta_q)  J_1(\rho \,q_{\alpha} \sin \theta_q)\:.
\end{multline}
The simplified expression for the spin-relaxation rate is
\begin{equation}
\begin{aligned}
\frac{1}{T_1}&=F_{ph} \frac{\nu_0^2 e^2 B^2}{2^{9} \pi \hbar  \rho} \int_0^\pi \, d\theta{_{ q}} \, \sin^3 \theta{_q} \cdot\\
& \left[ \sin^2 \theta{_q} \frac{q_l^3}{s_l^2} |\kappa_l|^2+(1+\cos^2 \theta{_q}) \frac{q_t^3}{s_t^2} |\kappa_t|^2 \right]\:,
\end{aligned}
\end{equation}
which can be calculated numerically using standard procedures.

\subsection{Analytic calculation of direct spin-phonon spin-relaxation rate}

To derive an analytical result we use trial wavefunctions of a Gaussian or exponential form. First, we approximate the ground state wave function by a Gaussian
\begin{equation}
\label{approx:gauss}
    \psi_{{\text{1s}}} = \frac{1}{(\sqrt{\pi}l)^{3/2}} e^{-{r^2}/{(2 l^2)}},
\end{equation}
with  $l=[1/(a_B^*)^2+1/{(2l_b^2)}]^{-1/2}$. The matrix element can be found analytically,
\begin{eqnarray}
\langle \text{1s} | \exp(i \mathbf{q r}) r_j | \text{1s} \rangle =  \frac{1}{2}i q_j l^2 e^{-q^2 l^2/4}\:,
\end{eqnarray}
where $j = x,y$.

Using Eqs.~\eqref{eq:T1start}, \eqref{eq:app_M_dir} we obtain the relaxation rate assuming that the spin-up state has a higher energy as compared with the spin-down one
\begin{eqnarray}
\label{gamma:dir}
    \Gamma_{\downarrow \uparrow} &=&  \frac{ (N_{ph}+1) v_0^2}{256 \pi \rho \hbar}  (e B l^2)^2 \sum_\alpha  \frac{(g \mu B)^5}{\hbar^5 s_\alpha^7} I_\alpha e^{-{q_{\alpha}^2 l^2}/{2}}, \qquad \\
    I_\alpha &=& \left \langle \left( \hat{e}_{x}^{\mathbf q, \alpha} \xi_y +  \hat{e}_{y}^{\mathbf q, \alpha} \xi_x \right)^2 \xi_x^2 \right \rangle_\Omega\:.
\end{eqnarray}
The integrals over phonon angle for the longitudinal and both transverse modes are
\({I_l = 4/35}\) and \({I_t = 16/105}\).
Taking into account the phonon factor $F_{ph}$, the final result for the relaxation rate by the direct spin-phonon process is 
\begin{multline}
    \frac{1}{T_1} =\frac{1}{2240 \pi} \frac{(e v_0 l^2)^2 (g \mu)^5 B^7}{\rho \hbar^6} \times \\
    \left( \frac{e^{-{q_l^2 l^2}/{2}}}{s_l^7}+\frac{4 e^{-{q_t^2 l^2}/{2}}}{3 s_t^7} \right) F_{ph}\:.
    \label{eq:direct_T1}
\end{multline}


Another possible choice of wave function for the donor-bound electron is an exponential 
\begin{equation}
\label{approx:exp}
\psi_{\text{1s}} = \frac{1}{\sqrt{\pi l^3}} e^{-r/l}.
\end{equation}
For this wave function,
\begin{eqnarray}
\langle \text{1s} | \exp(i \mathbf{q r}) r_j | \text{1s} \rangle =   \frac{i\, l^2 q_j}{(1+q^2 l^2/4)^3}\:.
\end{eqnarray}
The relaxation rate using an exponential wave function is the same as Eq.~\eqref{eq:direct_T1} with $\exp{(-q_\alpha^2 l^2/2)}$ replaced by ${4/(1+ q_\alpha^2 l^2/4)^6}$, in agreement with Eq.~\eqref{eq:direct} of the main text.

We note that in the presence of a magnetic field the form of the donor-bound electron functions depends on the gauge, which calls for special care in evaluating the matrix elements in Eq.~\eqref{me:direct:gauge}. Particularly, Eqs.~\eqref{approx:gauss} and \eqref{approx:exp} are valid in the symmetric gauge. For instance, in the Landau gauge, where ${\mathbf{A}=(0,Bx,0)}$, Eqs.~\eqref{approx:gauss} and \eqref{approx:exp} acquire extra phase factors $\exp{[i eB xy/(2\hbar)]}$. Taking these phase factors into account one can readily check that Eqs.~\eqref{me:direct:gauge} and, correspondingly, Eqs.~\eqref{gamma:dir}, \eqref{eq:direct_T1} and \eqref{eq:direct} are gauge invariant.





















\bibliography{toddkarin,toddkarin_extra,xiayu,kaimei,ioffe}

\begin{thebibliography}{67}
\bibitem [{ref({\natexlab{a}})}]{ref:InPfootnote}%
  \BibitemOpen
  \href@noop {} {}\bibinfo {howpublished} {Samples were grown by metalorganic
  vapour phase epitaxy at between 540 and 580$^\cdot$~C using trimethylindium
  and either phosphine or tertiarybutylphosphine.} ({\natexlab{a}})\BibitemShut
  {NoStop}%
\bibitem [{ref({\natexlab{b}})}]{ref:CdTefootnote}%
  \BibitemOpen
  \href@noop {} {}\bibinfo {howpublished} {The carrier concentration was not
  measured but is believed to be greater than $10^{14}$~cm$^{-3}$ but less than
  10$^{15}$~cm$^{-3}$ based on measurements on samples grown from the same
  starting material. The lineshapes of the one LO-phonon replica of the free
  exciton support this; the inverse Maxwell-Boltzmann distribution of {CdTe}-1
  is sharper than that of {CdTe}-2.} ({\natexlab{b}})\BibitemShut {NoStop}%
\end{thebibliography}

\end{document}